\documentclass[aps,reprint,twocolumn,superscriptaddress,citeautoscript,amsmath,amssymb,floatfix,longbibliography,prl]{revtex4-2}
\usepackage{graphicx}% Include figure files
\usepackage{times}
\usepackage{bm}% bold math
\usepackage{color}
\usepackage{gensymb}
\usepackage{dcolumn}% Align table columns on decimal point
\usepackage{xfrac}
\usepackage{braket}
\usepackage{siunitx}

\usepackage{soul}
\usepackage[resetlabels]{multibib}

\newcolumntype{d}[1]{D{.}{.}{#1}}
\begin{document}

\title{Hidden symmetry-breaking in a kagome Ising ferromagnet}
    \author{Tianxiong~Han}
	\affiliation{Ames National Laboratory, U.S. DOE, Iowa State University, Ames, Iowa 50011, USA}
	\affiliation{Department of Physics and Astronomy, Iowa State University, Ames, Iowa 50011, USA}
    \author{Tyler J. Slade}
	\affiliation{Ames National Laboratory, U.S. DOE, Iowa State University, Ames, Iowa 50011, USA}
    \author{Liqin Ke}
	\affiliation{Ames National Laboratory, U.S. DOE, Iowa State University, Ames, Iowa 50011, USA}
    \affiliation{Department of Materials Science and Engineering, University of Virginia, Charlottesville, VA 22904}
    \author{Qing-Ping Ding}
    \affiliation{Ames National Laboratory, U.S. DOE, Iowa State University, Ames, Iowa 50011, USA}
    \author{Minseong Lee}
        \affiliation{ National High Magnetic Field Laboratory, Los Alamos National Laboratory, Los Alamos, New Mexico 87545, USA}
    \author{R. D. McKenzie}
        \affiliation{Ames National Laboratory, U.S. DOE, Iowa State University, Ames, Iowa 50011, USA}
        \affiliation{Department of Physics and Astronomy, Iowa State University, Ames, Iowa 50011, USA}
    \author{Bing~Li}
	\affiliation{Ames National Laboratory, U.S. DOE, Iowa State University, Ames, Iowa 50011, USA}
        \affiliation{Neutron Scattering Division, Oak Ridge National Laboratory, Oak Ridge, TN, 37831, USA}
    \author{Dhurba R. Jaishi}
	\affiliation{Ames National Laboratory, U.S. DOE, Iowa State University, Ames, Iowa 50011, USA}
	\affiliation{Department of Physics and Astronomy, Iowa State University, Ames, Iowa 50011, USA}
    \author{Yongbin Lee}
	\affiliation{Ames National Laboratory, U.S. DOE, Iowa State University, Ames, Iowa 50011, USA}
    \author{D. M. Pajerowski}
        \affiliation{Neutron Scattering Division, Oak Ridge National Laboratory, Oak Ridge, TN, 37831, USA}
    \author{Qiang Zhang}
        \affiliation{Neutron Scattering Division, Oak Ridge National Laboratory, Oak Ridge, TN, 37831, USA}
    \author{Tao Hong}
        \affiliation{Neutron Scattering Division, Oak Ridge National Laboratory, Oak Ridge, TN, 37831, USA}
    \author{P. C. Canfield}
	\affiliation{Ames National Laboratory, U.S. DOE, Iowa State University, Ames, Iowa 50011, USA}
	\affiliation{Department of Physics and Astronomy, Iowa State University, Ames, Iowa 50011, USA}
    \author{Y. Furukawa}
	\affiliation{Ames National Laboratory, U.S. DOE, Iowa State University, Ames, Iowa 50011, USA}
	\affiliation{Department of Physics and Astronomy, Iowa State University, Ames, Iowa 50011, USA}
    \author{K. Thirunavukkuarasu}
	\affiliation{Florida A\&M University, Tallahassee, FL 32307 USA}
    \author{Aashish Sapkota}
	\affiliation{Ames National Laboratory, U.S. DOE, Iowa State University, Ames, Iowa 50011, USA}
    \author{Rebecca Flint}
	\affiliation{Ames National Laboratory, U.S. DOE, Iowa State University, Ames, Iowa 50011, USA}
	\affiliation{Department of Physics and Astronomy, Iowa State University, Ames, Iowa 50011, USA}
    \author{ R. J. McQueeney}
	\affiliation{Ames National Laboratory, U.S. DOE, Iowa State University, Ames, Iowa 50011, USA}
	\affiliation{Department of Physics and Astronomy, Iowa State University, Ames, Iowa 50011, USA}
\date{\today}

\begin{abstract}
Kagome metals can host unconventional electronic phenomena that emerge from their frustrated lattice geometry and associated band topology. Correlated electronic orders, such as charge-density waves and superconductivity, are observed to intertwine with subtle time-reversal symmetry breaking whose microscopic origin is not currently understood. Here, we provide evidence for such time-reversal symmetry breaking in the kagome metal TbV$_6$Sn$_6$ arising from staggered magnetic moments within the kagome layers. TbV$_6$Sn$_6$ consists of metallic V kagome layers separated by Tb triangular layers that host Ising ferromagnetic order. Deep in the ferromagnetic state, the Tb Ising doublet ground state should display a single, dispersionless spin-flip excitation. Instead, inelastic neutron scattering reveals two sharp excitations associated with inequivalent Tb sites, demonstrating that a symmetry-broken phase coexists with Ising ferromagnetism.
No additional structural or magnetic phase transitions are detected, and first-principles calculations rule out lattice distortions as the origin of the splitting.  We attribute this effect to time-reversal symmetry breaking encoded by small V moments that couple to the Tb sublattice and leave a measurable spectral fingerprint. Our results establish rare-earth local moment spectroscopy as a sensitive probe of subtle broken symmetries and highlight an unexpected interplay between kagome magnetism and rare-earth local moment magnetism.
\end{abstract}

\maketitle
{\it Introduction}. Kagome metals are widely studied for their capacity to host a variety of correlation-driven electronic states that arise from special features of their band structure \cite{Jiang:2021aa, Teng:2022aa, PhysRevLett.127.046401, PhysRevMaterials.5.034801, PhysRevMaterials.3.094407, PhysRevB.106.184422, PhysRevB.108.045132,Yin:2020aa, Baidya:2019aa}. Recently, there has been great interest in vanadium-based $A$V$_3$Sb$_5$ kagome metals following the observation of charge-density wave (CDW) order  \cite{PhysRevX.11.031026, PhysRevLett.126.247001,PhysRevB.103.L220504,PhysRevX.11.031050,graham2024depth}. This CDW state breaks translational symmetry and is also reported to simultaneously break time-reversal symmetry, presumably due to loop currents that form in concert with complex charge ordering \cite{ PhysRevLett.127.217601, mielke2022time}.  In this unique quantum state, the appearance of chiral superconductivity is a surprising development that provides a test bed to study the interrelationship between various correlation-driven symmetry-broken states \cite{FENG20211384, PhysRevB.85.144402}.

Beyond the $A$V$_3$Sb$_5$ family, other V-based kagome metals with distinct structural frameworks have also shown intriguing symmetry-breaking behavior. A CDW symmetry-breaking phase transition has recently been discovered in ScV$_6$Sn$_6$ from a different mechanism than that in the $A$V$_3$Sb$_5$ compounds \cite{PhysRevLett.129.216402,lee2024nature}. Muon spin resonance data suggest that time-reversal symmetry may also be broken in the CDW state \cite{guguchia2023hidden}. Other $R$V$_6$Sn$_6$ compounds readily admit magnetic rare-earth ions ($R={\rm Gd}-{\rm Tm}$), allowing for the interplay of local moment magnetism with symmetry-broken metallic V kagome layers.  Unfortunately, while the rare-earth moments in $R$V$_6$Sn$_6$ compounds order at low temperatures ($T_{\rm C}<5$ K) with transport anomalies highlighting the coupling to conduction electrons \cite{PhysRevResearch.6.043291, PhysRevB.106.115139, PhysRevMaterials.6.104202, PhysRevB.109.104412, PhysRevMaterials.6.105001}, no other symmetry-broken phases, such as CDW order, have been detected. However, the observation of large orbital magnetization in weak magnetic fields in TbV$_6$Sn$_6$ suggests that novel magnetic interplay with kagome electrons is possible \cite{li2024spin}.

Here, we investigate the magnetic interactions in TbV$_6$Sn$_6$ using inelastic neutron scattering (INS).  TbV$_6$Sn$_6$ is a ferromagnet (FM) [$T_{\rm C}=4.4$ K] with strong Ising anisotropy where Tb ions adopt a high-spin electronic doublet ground state \cite{PhysRevMaterials.6.104202, PhysRevB.106.115139,han2024proximity}. In the FM state, this doublet is split by the molecular field of the surrounding Tb ions, and INS should detect a single spin-flip excitation between these two states. However, we observe two spin-flip excitations at different energies, rather than one, and conclude that symmetry-breaking is present which leads to two inequivalent Tb sites.

Due to the non-Kramers nature of the Tb ion ($J=6$), both crystallographic and time-reversal symmetry breaking can generate different crystalline-electric field (CEF) or molecular field magnitudes, respectively, that locally shift the spin-flip excitation energy. Our analysis of the spin-flip excitations and their temperature dependence indicates a presence of a staggered internal field that breaks translational and time-reversal symmetry well above $T_{\rm C}$. This hidden symmetry breaking is distinct from Tb ordering, with the most likely origin being small-moment antiferromagnetism that develops within the V kagome layers.

{\it Magnetic Hamiltonian for TbV$_6$Sn$_6$}. The Tb ions in TbV$_6$Sn$_6$ have large angular momentum ($J=6$, $S=3$, $L=3$, $g_J=3/2$) and are located on a triangular lattice with $6/mmm$ point group symmetry.  The CEF Hamiltonian is  $\mathcal{H}_{CEF}=B_2^0\mathcal{O}_2^0+B_4^0\mathcal{O}_4^0+B_6^0\mathcal{O}_6^0+B_6^6\mathcal{O}_6^6$, where $B_l^m$ are CEF parameters and $\mathcal{O}_l^m$ are Stevens operators \cite{cef_point}. The $B_6^6$ CEF parameter ultimately plays an important role, but in its absence all zero-field CEF states can be labeled by the azimuthal quantum number,  $\ket{J,m} = \ket{m}$. Previous work \cite{han2024proximity} finds a $\ket{m=\pm6}$ doublet ground state that is separated from the first excited state $\ket{\pm5}$ by a large CEF energy of $\Delta \approx 8$ meV $\gg k_{\rm B}T_{\rm C}$, consistent with Ising character.

Tb moments ${\bf J}$ within the triangular layer are magnetically coupled through a combination of competing FM RKKY exchange and AFM dipolar couplings \cite{PhysRevResearch.6.043291, PhysRevB.108.035134}. As described in the SI, the effective nearest-neighbor coupling is $\mathcal{J} \approx \mathcal{J}_{RKKY} + \mathcal{J}_{dipolar}$ where $\mathcal{J}_{dipolar}/\mathcal{J}_{RKKY} \approx -0.20$.  Single-crystal INS data find that the FM interlayer coupling $\mathcal{J}_z$ is very weak and likely dipolar in origin, resulting in a quasi-2D Ising FM.

\begin{figure}[]      
    \centering 
	\includegraphics[width=0.95\linewidth]{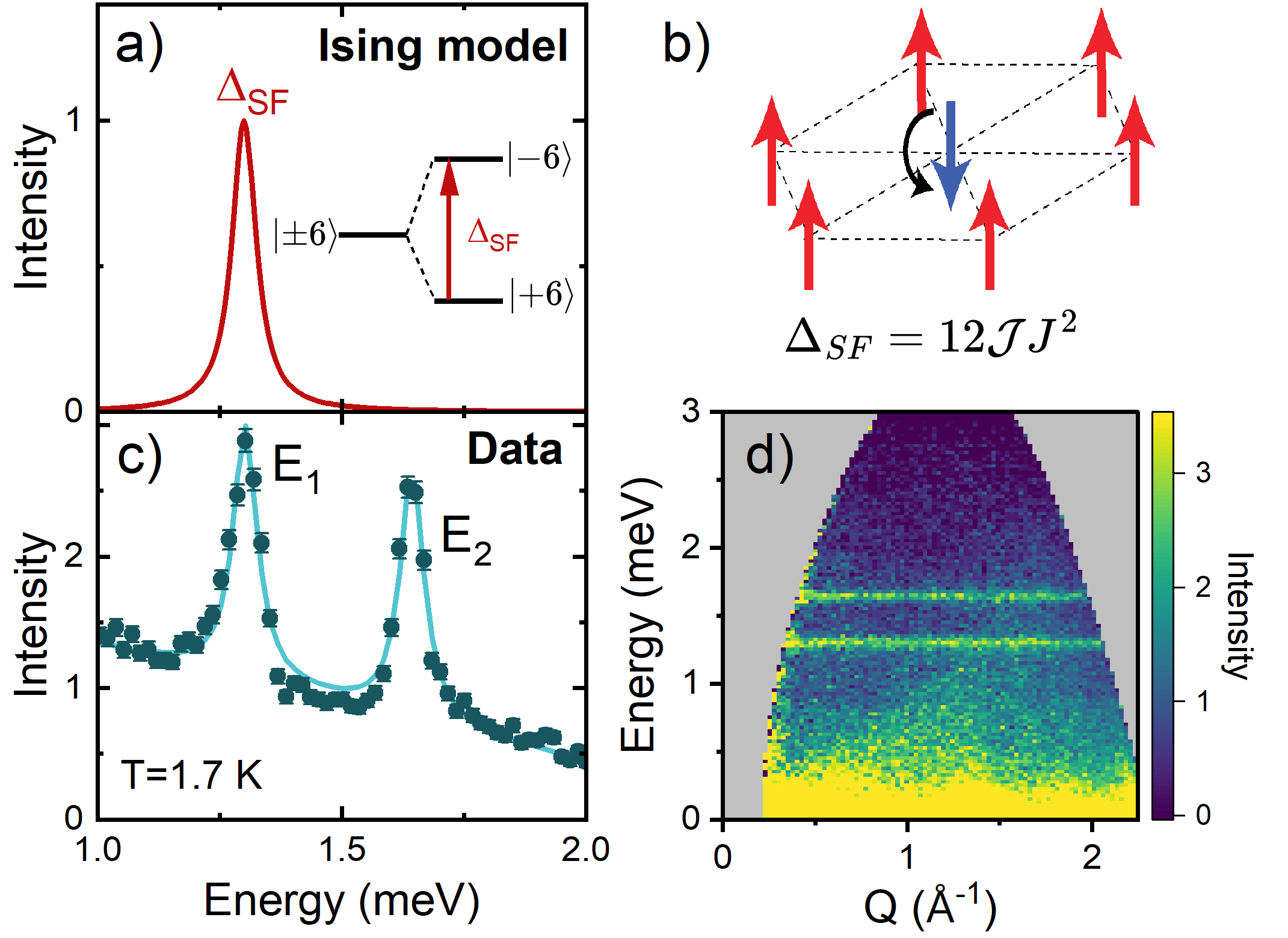}
	\caption{\textbf{Spin-flip excitations in an Ising ferromagnet.} (a) Expected spin-flip excitation spectrum within the $\ket{\pm 6}$ doublet in the FM ordered state. The spectrum should consist of a single peak at $\Delta_{SF}\approx 1.3$ meV. (b) Schematic diagram of the spin-flip excitation. (c) Inelastic neutron scattering data from TbV$_6$Sn$_6$ at $T=1.7$ K showing two spin-flip excitations labeled $E_1$ and $E_2$. (d) Full $Q-E$ neutron spectrum of TbV$_6$Sn$_6$ at 1.7 K.}
 \label{fig:simple_Ising}
\end{figure}

{\it Expected excitation spectrum of an Ising FM}. The strong uniaxial anisotropy of an Ising FM means that the magnetic moments can only point up or down. From the perspective of quantum states in TbV$_6$Sn$_6$, the up and down moments are represented by the $\ket{\pm 6}$ doublet ground state where the spin-flip transition from $\ket{+6}\rightarrow \ket{-6}$ is the characteristic excitation of an Ising FM. The spin-flip energy is the cost to flip one spin while all neighboring spins remain mutually parallel and is given by $\Delta_{SF} = z|\mathcal{J}|J\Delta m = 12|\mathcal{J}|J^2$ where $z=6$ is the coordination number for triangular lattice and $\Delta m=2J$, as shown in Fig.~\ref{fig:simple_Ising}(a) and (b). 

The spin-flip energy can be estimated from $T_{\rm C}$ by using the relation $k_{\rm B}T_{\rm C}= [4/\ln(3)]|\mathcal{J}|J^2$ for a 2D Ising FM \cite{newell1950crystal}. We find that $\mathcal{J} \approx  2.9~\mu$eV and $\Delta_{SF} \approx 1.3$ meV. Within a mean-field approximation, $\Delta_{SF}=2\beta J$ is caused by a molecular field of strength $\beta=z|\mathcal{J}|J \approx  0.10$ meV [or $B_{\text{MF}} = \beta/(g_J\mu_{\rm B})=$ 1.2 T]. Single-crystal INS data measuring the dispersion of the main CEF transition from $\ket{+6}\rightarrow\ket{+5}$ provide a consistent independent measure of $\mathcal{J}$ (see SI).

Since only dipole-allowed transitions with $\Delta m = 0, \pm 1$ are detectable with INS \cite{BoothroydBook}, the spin-flip transition with $\Delta m=12$ should not be observed. However, the spin-flip transition becomes weakly dipole-allowed when $B_6^6\mathcal{O}_6^6 \propto J_{\pm}^6\neq0$ since level mixing among $\ket{\pm 6}$ and $\ket{0}$ states opens a channel with $\Delta m = 0$. As discussed below, a non-zero $B_6^6$ will also split the non-Kramers $\ket{\pm 6}$ Ising doublet by $\lesssim$ 0.01 meV which is much smaller than the spin-flip energy.

{\it Inelastic neutron scattering data}. INS data reveal a different picture of the spin-flip excitations in TbV$_6$Sn$_6$ than that described above. We find {\it two} transitions (labeled $E_1$ and $E_2$) close to the estimated spin-flip energy of 1.3 meV, as shown in Fig.~\ref{fig:simple_Ising}(c). At temperatures well below $T_{\rm C}$, Figs.~\ref{fig:simple_Ising}(c) and (d) show that the excitations are sharp (resolution-limited), dispersionless, and have equal intensities.  Given these traits and the proximity to the estimated $\Delta_{SF}$, we identify both peaks as spin-flip excitations with a relative shift of $2\delta=E_2-E_1=0.34$ meV. 

\begin{figure*}[]      
    \centering 
	\includegraphics[width=.85\linewidth]{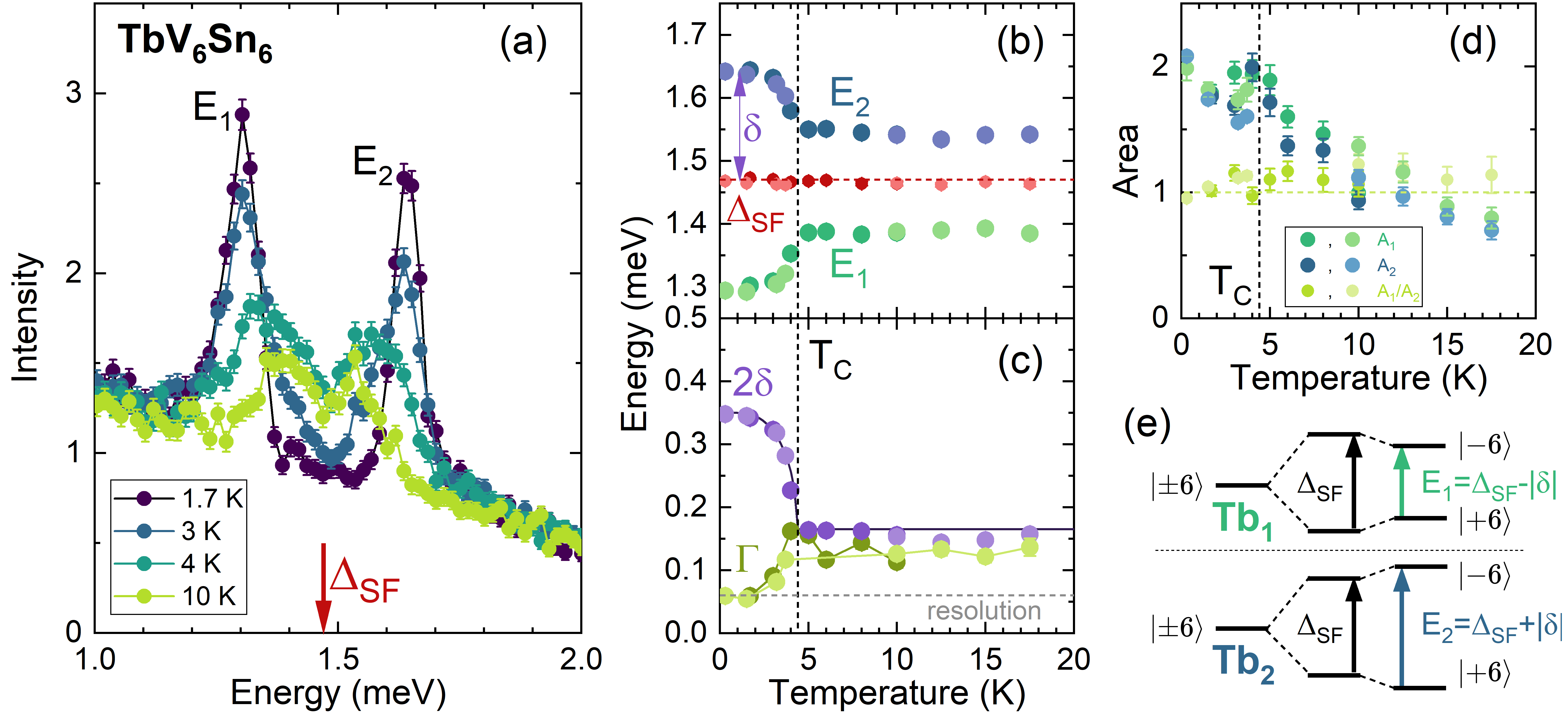}
	\caption{\textbf{Temperature evolution of spin-flip excitations in TbV$_6$Sn$_6$.} (a) Inelastic neutron scattering data showing two spin-flip excitations at energies of $E_1$ and $E_2$ at temperatures above and below $T_\text{C}$. (b) Temperature dependence of $E_1$ and $E_2$, average value ($\Delta_{SF}$), and splitting parameter ($2\delta=E_2-E_1$) as obtained from fitting INS spectra similar to panel (a). (c) Temperature dependence of $2 \delta$ and Lorentzian full-width $\Gamma$ of the excitations. (d) The temperature dependence of the area of each peak along with the ratio of the areas $A_1/A_2$. (e) Schematic level diagrams for two inequivalent Tb ions with different spin-flip energies caused by symmetry-breaking energy $\delta$.  In panels (b)-(d), different shading of symbols indicates independent measurements of different samples.}
 \label{fig:splitting}
\end{figure*}

When the temperature is increased, Figs.~\ref{fig:splitting}(a)$-$(c) show that $2\delta$ is reduced and the Lorentzian full-width ($\Gamma$) broadens until $T_{\rm C}$ is reached. $E_1$ and $E_2$ have a component that is proportional to the FM order parameter below $T_{\rm C}$, but with opposite sign ($\pm \delta$). Surprisingly, the spin-flip excitations do not vanish above $T_{\rm C}$, maintaining a constant energy shift of $2\delta= 0.16$ meV until at least 17.5 K. The average of $E_1$ and $E_2$ is temperature-independent and we identify it as the experimentally determined spin-flip energy $\Delta_{SF}=1.47$ meV (slightly larger than estimates from $T_{\rm C}$ noted above). Fig.~\ref{fig:splitting}(d) shows that the intensity of the two peaks weaken with increased temperature, but always maintain equal relative intensities.

Presumably, the persistence of spin-flip excitations above $T_{\rm C}$ is due to strong short-range FM correlations that are observed to persist up to at least 40 K (See SI Fig. S6). Unlike mean-field theory where the spin correlations and the molecular field vanish above $T_{\rm C}$, exact solutions of the Ising model show that robust nearest-neighbor spin correlations maintain a quantized value of the local field and corresponding spin-flip energy above $T_{\rm C}$ \cite{Thomsen84}. 

Thermal fluctuations will eventually suppress the spin-flip excitations at high enough temperature. Unfortunately, the spin-flip signals are overwhelmed above 20 K by the growth of other CEF excitations (see Ref.~\cite{han2024proximity} for details). In this respect, we note that the spin-flip excitations are very weak, with intensities more than $400$ times weaker than the main dipole-allowed $\ket{+6}\rightarrow\ket{+5}$ CEF transition out of the ground state. This weakness is expected due to the small dipole matrix element between the split doublet states, as noted above.

%\subsection{Analysis}
{\it Analysis}. We consider several possibilities to explain the presence of two spin-flip excitations in TbV$_6$Sn$_6$. One possibility is that hyperfine interactions of the form $AJ_zI_z$ lead to a splitting between coupled nuclear and electronic spin states.  Terbium has a nuclear spin of $I=3/2$ and the estimated hyperfine coupling constant $A = 0.002$ meV, as reported in Refs.~\cite{Bleaney_1961,Baker_1955}, is consistent with the nuclear Schottky anomaly observed in TbV$_6$Sn$_6$ \cite{PhysRevResearch.6.043291}. Hyperfine coupling will result in four electronic spin dipole-allowed transitions, not two, and the overall splitting of $36A \approx 0.07$ meV is much smaller than $2\delta$ and comparable to the instrumental resolution (0.06 meV full-width-at-half-maximum). Nuclear quadrupolar interactions are expected to be even smaller \cite{kotzler1980effect}, and so we dismiss coupling to nuclear spins as a viable explanation.

%We might also consider some mechanism for quenched disorder that creates different local Tb environments. However, this would lead to inhomogeneous broadening and not two sharp peaks. Furthermore, the high quality of our x-ray diffraction data as shown in SI does not provide evidence for antisite mixing or vacancies.

The sharpness and equal intensity of the two peaks are more consistent with a scenario where a symmetry-breaking transition occurs which generates two inequivalent Tb sites (Tb$_{1}$ and Tb$_{2}$) with different CEFs or molecular exchange fields, as shown in Fig.~\ref{fig:splitting}(e). Given the subtle broken-symmetry phases found in other kagome metals, here we consider the possibility that either crystallographic or time-reversal symmetry is broken at temperatures well above $T_{\rm C}$, generating a symmetry-breaking field acting on the Tb ion with strength $\delta$.

\begin{figure*}[]      
    \centering 
	\includegraphics[width=.85\linewidth]{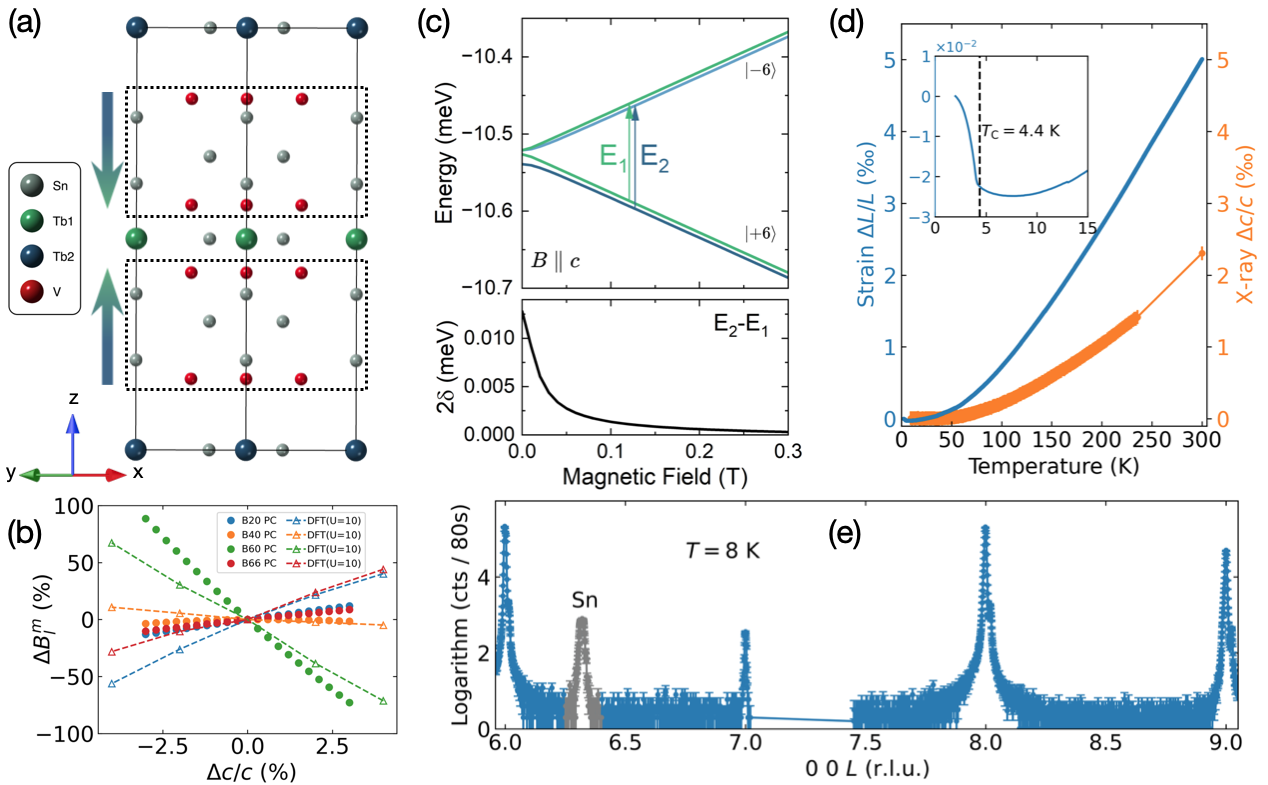}
	\caption{\textbf{Crystallographic-symmetry breaking.} (a) A possible crystallographic-symmetry breaking scenario showing inequivalent Tb sites with equal multiplicities caused by $(0,0,1/2)$ distortion. The displacement of the atoms in dashed boxes creates inequivalent Tb$_1$ (light blue atoms) and Tb$_2$ (dark blue) local environments with Tb-Sn layer distances $d_0\pm\epsilon$ and can be understood as a local $\Delta~c/c$ distortion. 
    (b) The change of $B_{l}^m$ parameters with  $\Delta~c/c$ lattice distortion using the point charge model (dots) and the DFT calculation (triangle with dashed lines) up to $\Delta~c/c = 4\%$.
    (c) Evolution of the doublet splittings $E_1$ and $E_2$ in a molecular field along the $c$-axis in the presence of a large local distortion ($\Delta c/c  = 3\%$) that generates Tb$_1$ and Tb$_2$ ions with $B_6^6$ values shifted $\pm 30\%$.  The lower panel shows that the zero-field shift $2\delta < 0.015$ meV is suppressed by the molecular field.
    (d) Temperature-dependent strain ($\Delta L/L$) and X-ray diffraction measurements ($\Delta c/c$). The inset shows a spontaneous magnetostriction of $\Delta L/L \sim 0.002\%$ is observed below $T_\text{C}$.
    (e) Logarithm of XRD intensity along $(0~0~L)$ direction at $8$~K. No half-integer $L$ peaks are observed that correspond to ${\bf q}=(0,0,1/2)$ symmetry breaking. The Bragg peak corresponding to the Sn impurity is colored in gray. The shoulders next to the Bragg peaks are from Cu-K$\alpha_2$ incident X-ray.
    % 
    %(d) The percentage variance of B$_{l}^m$ versus the variance of lattice distortion by changing volume calculated from the point charge model. The B$_6^6$ varies less than 10\% with the lattice distorted by 1\% in either case. 
    %.
    }
 \label{fig:CSbreaking}
\end{figure*}

{\it Crystallographic symmetry breaking}. We first consider crystallographic symmetry breaking which can lead to site-to-site variation of the spin-flip energy due the non-Kramers' nature of the Tb ion.  At high temperatures, all $R$V$_6$Sn$_6$ compounds adopt the $P6/mmm$ space group with Tb located on the 1$a$ Wyckoff site, i.e. all Tb sites are crystallographically equivalent. ScV$_6$Sn$_6$ is known to undergo a structural transition below 92 K that would generate inequivalent Sc sites \cite{PhysRevLett.129.216402,lee2024nature}. The CDW consists of trimerization of Sn-Sc-Sn units along the $c$-axis that are staggered in the plane. A distortion at ${\bf q}_{\text{CDW}}=(1/3,1/3,1/3)$ creates two inequivalent Tb sites with multiplicity ratio of 2:1, which is inconsistent with the equal intensity of the spin-flip excitations. Various experimental studies have not detected a structural distortion in any $R$V$_6$Sn$_6$ compound other than $R=$ Sc \cite{PhysRevResearch.6.043291, PhysRevB.104.235139, ishikawa2021gdv6sn6, PhysRevMaterials.6.104202, PhysRevB.106.115139}.  Our temperature-dependent strain, single-crystal X-ray diffraction, powder neutron diffraction, nuclear magnetic resonance, and Raman scattering experiments (Fig.~\ref{fig:CSbreaking}(d) and SI) find no evidence of CDW order or other structural transitions down to 2 K.

However, it is possible that subtle (not yet detected) crystallographic symmetry breaking is present or may occur close to $T_{\rm C}$ due to magnetoelastic interactions. ISODISTORT \cite{campbell2006isodisplace} predicts only one maximal subgroup which generates an equal population of unique Tb$_1$ and Tb$_2$ sites. This distortion with ${\bf q}=(0,0,1/2)$ is caused by the dimerization of Sn or V layers along $c$, as shown in Fig.~\ref{fig:CSbreaking}(a). This generates local $\Delta c/c$ distortions with different $B_6^6$ values which leads to different ground state doublet splittings for Tb$_1$ and Tb$_2$.

 In Fig.~\ref{fig:CSbreaking}(b), calculations using point-charge models provided by $PyCrystalField$ \cite{Scheie:in5044} and density-functional theory $+U$ (DFT$+U$) predict changes in the $B_l^m$ parameters when varying $\Delta c/c$. Fig.~\ref{fig:CSbreaking}(c) shows that the spin-flip energies of the Tb$_1$ and Tb$_2$ sites are only slightly shifted ($\delta < 0.01$ meV) by a large strain of $\Delta c/c = 3$\% (corresponding to a $\sim30\%$ change in $B_6^6$), and $\delta$ is further suppressed by the molecular field.  However, temperature-dependent strain measurements shown in the inset to Fig.~\ref{fig:CSbreaking}(d) find a tiny spontaneous magnetostriction of $\Delta L/L < 0.003\%$ below $T_{\rm C}$. Upon warming up to $300$~K, the lattice expands less than $0.5\%$ from both temperature-dependent strain and X-ray measurements. The differences between strain and X-ray results may come from the stacking of multiple pieces when measuring strain.  X-ray diffraction data shown in Fig.~\ref{fig:CSbreaking}(e) also find no evidence for (0,0,1/2)-type crystallographic symmetry lowering. While we cannot entirely dismiss crystallographic symmetry breaking as a source for $\delta$ for ions without Kramers degeneracy, we estimate that these effects are orders-of-magnitude too small to account for the observed splitting shown in Fig.~\ref{fig:splitting}.

\begin{figure}[]      
    \centering 
	\includegraphics[width=1\linewidth]{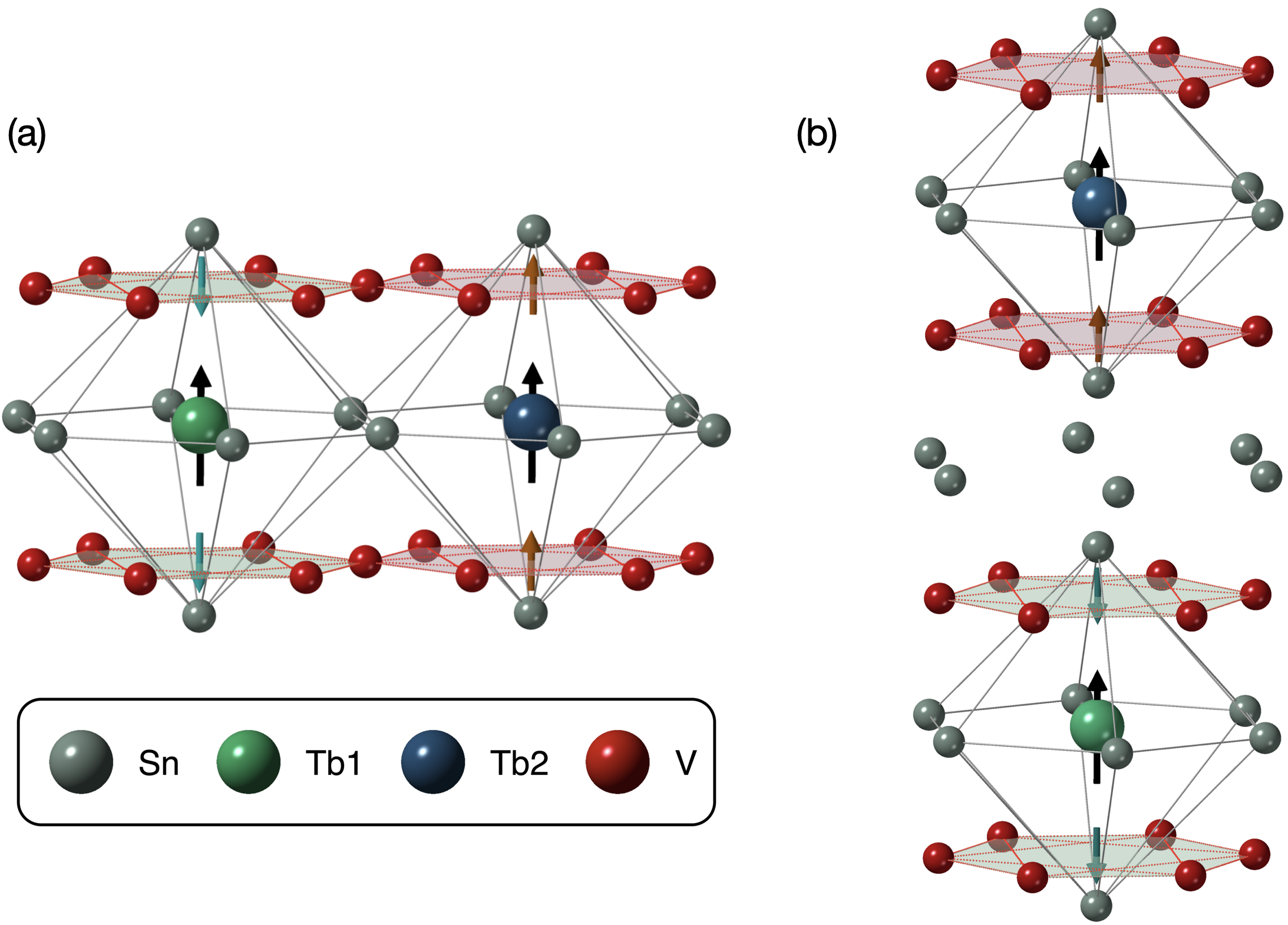}
	\caption{\textbf{Time-reversal symmetry breaking.} Different magnetic environments of Tb$_1$ and Tb$_2$ ions due to (a) stripe-like antiferromagnetic order within the V kagome layer and (b) up-up-down-down AFM stacking of FM V layers.}
 \label{fig:TRSbreaking}
\end{figure}

{\it Time-reversal symmetry breaking}. We now consider the possibility that an interaction exists with the Tb sublattice which breaks time-reversal symmetry. Here, we speak of time-reversal symmetry breaking that is distinct from FM ordering of the Tb ions and presumably arises from the V layers. To deliver two different spin-flip energies, this must take the form of a staggered field that also breaks translational symmetry (i.e. antiferromagnetic) as shown in Fig.~\ref{fig:TRSbreaking}. 

However, no such staggered magnetism has been directly detected by neutron diffraction. We investigated potential V magnetism using $^{51}$V nuclear magnetic resonance (NMR) measurements on single-crystal and polycrystalline samples of TbV$_6$Sn$_6$ performed at a Larmor field of 7 T, as described in the SI. In the FM ordered phase, the NMR resonance shifts away from the Larmor field by 0.39 T, but no splitting is observed that would indicate AFM order. By accounting for several contributions to the shift, including macroscopic and dipolar fields, we cannot rule out the existence of a small V moment $\sim 0.02~\mu_B$. This suggests tiny itinerant V moments may exist, perhaps originating from orbital magnetization similar to that reported in Ref.~\cite{li2024spin}.

We hypothesize that an intersublattice magnetic coupling exists of the form $\mathcal{J}^{TV} {\bf J} \cdot {\bf s}$ that couples the Tb moment (${\bf J}$) to a staggered spin degree-of-freedom associated with the V layers ({\bf s}). To account for the splitting observed above $T_{\rm C}$, the V AFM order must set in above $T_{\rm C}$ with some unknown exchange energy scale, $T_N \sim \mathcal{J}^{VV}s^2$.  The V magnetic configuration cannot be directly determined from INS data, but we discuss some possibilities below.  

In Fig.~\ref{fig:TRSbreaking}(a), we show two unique Tb sites that appear between stripe-like AFM V layers with FM V-V interlayer coupling. Tb$_2$ resides in a V generated field of $\delta = 12Js\mathcal{J}^{TV}$ and Tb$_1$ in a field of $-\delta$. The INS data provide an estimate of $s\mathcal{J}^{TV}\approx0.002$ meV which is significantly smaller than that found in ferrimagnetic TbMn$_6$Sn$_6$ ($s\mathcal{J}^{TM}=1.8$ meV) with large Mn moments \cite{riberolles2023orbital}. A key constraint for this scenario is that the Tb sublattice must remain FM in the presence of AFM V order. Therefore, there must be an energy barrier that prevents the Tb$_1$ site from flipping parallel to the V moments. This condition $\Delta_{SF} > 3\delta/2$ is easily satisfied and guarantees that two spin-flip excitations will be observed.

Fig.~\ref{fig:TRSbreaking}(b) shows a scenario where FM V layers are staggered in an up-up-down-down configuration with ${\bf q} = (0,0,1/2)$, leading to inequivalent Tb$_1$ and Tb$_2$ sites in a similar fashion to the crystallographic distortion shown in Fig.~\ref{fig:CSbreaking}(a). In this scenario, FM order on the Tb sublattice must be maintained by Tb-Tb interlayer coupling ($\mathcal{J}_z$) such that $\mathcal{J}_zJ^2>\delta/2$, but the dispersionless character of the main CEF excitation perpendicular to the layer provides an upper limit of $\mathcal{J}_z < 0.0015$~meV. Thus, the scenario in Fig.~\ref{fig:TRSbreaking}(b) appears unfeasible as it interactions will favor AFM Tb order.

The temperature dependence of $\delta(T)$ shown in Fig.~\ref{fig:splitting}(c) can naturally be explained if the Tb and V order parameters have an attractive quadratic Landau coupling, $-\gamma M_{Tb}^2 N_{V}^2$, where $N_V$ is the V staggered magnetization.  This coupling would cause a change in the proposed V order parameter just below $T_{\rm C}$, $\Delta N_V(T) \propto M_{Tb}(T)^2$ that shifts the two spin-flip excitation energies in opposite directions. The fit of $\delta(T)$ versus the $(1~0~0)$ Bragg peak intensities, shown in the SI, agrees with $\delta(T)\propto M_{\text{Tb}}^2$ near the transition. By contrast, $\delta(T) \propto M_{\text{Tb}}$ would imply a ferromagnetic V order $M_V$, with a linear coupling, $\gamma M_{Tb} M_{V}$. This scenario cannot explain the two distinct Tb sites whose spin-flip excitations move in opposite directions below $T_{\text{C}}$, and the $\gamma = 0$ phase transition $T_{\rm C}$ would be smeared out by the effective field due to $M_V$ to a degree that is inconsistent with the specific heat data \cite{PhysRevMaterials.6.104202}.  Therefore, a stripe-like antiferromagnetic V ordering is the most consistent with our data.

{\it Discussion}. In TbV$_6$Sn$_6$, the strong Ising character of the ground state limits the possible magnetic excitations that can occur in the FM ordered phase. We have, in essence, the simplest possible quantum system consisting of a doublet that is split by its local environment. This will result in a single spin-flip excitation when all Tb sites are equivalent.  The observation of two spin-flip excitations can only arise when two inequivalent Tb ions reside in different local environments. The relative shift in the doublet splittings therefore acts as a very sensitive probe of the local symmetry-breaking and its strength.

Our analysis shows that the doublet splitting of Tb$_1$ or Tb$_2$ is dominated by the exchange field of the surrounding Tb ions ($\Delta_{SF}$) combined with a weaker symmetry-breaking interaction $\delta/\Delta_{SF} \approx 10\%$. Due to the non-Kramers nature of the Tb ion, this interaction can arise from either crystallographic or time-reversal symmetry breaking, although we argue that any crystallographic symmetry-breaking contributes  $\ll1\%$ to the local field. Thus, it is much more likely that Tb ions experience a staggered internal field that breaks both time-reversal and translational symmetries.

Outside of Tb itself, the origin of time-reversal symmetry breaking could only arise from spins or orbital motion of electrons within the V kagome layers. We hypothesize that magnetic V-V interactions drive AFM order that sets in well above $T_C = 4.4$ K, but V carries a small moment that so far has not been detected by neutron diffraction, but cannot be ruled out by NMR data. For local V spins, the development of robust AFM order within a kagome layer [Fig.~\ref{fig:TRSbreaking}(a)] may be hampered by geometric frustration. Another possibility is the development of orbital magnetization within the kagome layers. It has been shown that weak external magnetic fields can strongly split Dirac cones near $E_F$, resulting in spin Berry curvature that generates a momentum-dependent orbital magnetization which is peaked near the $K$-point in the Brillouin zone \cite{li2024spin}. However, it is not clear whether such orbital moments can spontaneously form and order.

In any case, the proposed coupling between V and Tb moments is small, with V providing an effective staggered magnetic field of 0.15 T acting on the Tb site. This is $\sim$1000 times weaker than the Mn molecular field experienced by Tb in ferrimagnetic TbMn$_6$Sn$_6$ \cite{PhysRevX.12.021043, riberolles2024chiral, PhysRevB.111.054410}. Other methods, such as muon spectroscopy, should be employed to provide additional evidence for potentially novel magnetism occurring within the kagome bands.

{\it{Summary}}.
We report the observation of two spin-flip excitations in the Ising FM TbV$_6$Sn$_6$ that originate from distinct Tb local environments. We hypothesize that the local environments are distinguished by small, staggered magnetic moments occurring on the V kagome sublattice. Although the V magnetic order has not been directly detected, it nevertheless breaks translational symmetry at the rare-earth sites and give rise staggered molecular field and a measurable splitting in the rare-earth spin excitation spectrum. These findings indicate a novel time-reversal symmetry breaking in the kagome electrons and establish a new method for detecting subtle symmetry breaking in quantum materials.

%This scenario suggests either itinerant behavior or the emergence of novel kagome magnetism, such as a spontaneous ordering of orbital magnetization associated with Berry curvature in this topological material.

\section{Methods}
{\it Crystal growth and characterization}.
Single crystals of TbV$_6$Sn$_6$ were synthesized using the Sn flux method. High-purity elemental Tb (Ames Lab, 99.99\%), V (Alpha-Aesar, 99.99\%), and Sn (Alpha-Aesar, 99.99\%) were combined in a molar ratio of 1:6:60 and placed into the growth chamber of a 5 ml alumina Canfield crucible set \cite{canfield2016use}. The crucibles were sealed under vacuum in a fused silica ampule and heated in a box furnace to \SI{1180}{\celsius}. After holding at \SI{1180}{\celsius} for $\approx$ 24 h, the samples were removed from the furnace and the liquid decanted in a metal centrifuge, upon which the liquid phase is cleanly captured in the second crucible on the top side of the crucible set. 
This first step allows us to cleanly separate the liquid from any undissolved V \cite{slade2022use}.
A second sequence is then carried out, with the captured decant used in the ``growth" side of a new crucible set. The second crucible set was sealed under vacuum as before and warmed in a box furnace to \SI{1195}{\celsius}. After holding at \SI{1195}{\celsius} for 6 h, the furnace was cooled to \SI{775}{\celsius} over $\sim200$~h, upon which the ampule was removed from the furnace and the excess liquid decanted. Once cool, the ampules and crucibles were opened to reveal relatively large (10-30 mg), single-phase, crystals of TbV$_6$Sn$_6$.The polycrystalline sample was prepared by grinding the single crystal using a mortar and pestle, and sieving with a diameter of 32 micrometers to obtain particles of uniform size.

{\it Inelastic neutron scattering measurements}. INS measurements on polycrystalline and single-crystal samples of TbV$_6$Sn$_6$ were performed on the Cold Neutron Chopper Spectrometer (CNCS) at the Spallation Neutron Source at Oak Ridge National Laboratory using incident neutron energies of $E_{i} = 3.32$~meV and $12$~meV. The measured INS intensity is proportional to the spin-spin correlation function $\mathcal{S}(Q, \omega)$ with $Q$ being the momentum transfer, and $\omega = \Delta E / \hbar$ being the energy transfer. The data is histogrammed with bin sizes of $\delta E = 0.06$~meV and $\delta Q = 0.02$~\AA$^{-1}$ using $DAVE$ software \cite{azuah2009dave}. After integrating the intensity over the range $ 0.1\leq Q\leq2.7 $, the resulting spectrum is fit using a Lorentzian function, accounting for both the instrumental and intrinsic resolutions through convolution.

{\it Neutron Diffraction}.
We performed neutron powder diffraction measurements using the time-of-flight powder diffractometer POWGEN at Spallation Neutron Source at Oak Ridge National Laboratory. Approximately $1.3$ grams of samples were prepared after grinding single crystals to a fine powder and sieved with $32~\mu$m. Diffraction patterns were collected in the paramagnetic regime at $15$ K and the FM ordered state at $1.5$~K using a neutron beam with a center wavelength of $2.665~\AA$. Rietveld refinements using the FullProf suite \cite{rodriguez1993recent} were performed to determine the magnetic structure.

Neutron diffraction measurements were also performed on the powder sample of TbV$_6$Sn$_6$ using the cold neutron triple-axis spectrometer (CTAX) at High Flux Isotope Reactor (HFIR), ORNL, to get the order parameter. PG(002) monochromators and analyzers were used. Both the incident and final neutron energies were set to 3.5 meV. A cooled Be filter was placed after the sample table to remove the higher-order harmonics. 

{\it X-ray diffraction}. High-resolution single-crystal X-ray diffraction measurements were performed at Ames National Laboratory using a four-circle diffractometer with Cu-K$\alpha_1$ radiation from a rotating-anode source and a Ge (1, 1, 1) monochromator. The sample was attached to a flat Cu mount which was thermally anchored to the cold head of a He closed-cycle refrigerator. Be domes were used as vacuum shrouds and heat shields. A small amount of He exchange gas facilitated thermal equilibrium. The lattice constant $c$ was measured with $\theta-2\theta$ scans at $\textbf{Q}=(0,0,8)$ while warming from 8 K to 300K. Measurement along $(0,0,L)$ direction is performed at 8 K.

{\it Temperature-dependent strain}. The temperature-dependent strain was measured using the dilatometry option of the Quantum Design DynaCool system. The measurement principle and setup are described in detail by Martien et. al. \cite{dilatometry}. To enhance measurement sensitivity, five single crystals were stacked along the crystallographic $c$-axis. Since the total thickness was insufficient to fill the dilatometer gap, a FuSi spacer provided by Quantum Design was added to fill the remaining space; no adhesive was used in the assembly. The thermal dilation of the FuSi spacer was measured independently, and the final sample signal was obtained by subtracting the FuSi background from the combined FuSi-plus-sample measurement.

{\it Nuclear magnetic resonance}. The nuclear magnetic resonance (NMR) measurements of $^{51}$V nuclei ($I=\frac{7}{2},\frac{\gamma_\text{N}}{2\pi}=11.193$~MHz/T, $Q=-0.052$~barns) were conducted on single crystals and a loosely packed powder using a laboratory-built phase-coherent spin-echo pulse spectrometer by sweeping the magnetic field $B$ at fixed resonance frequencies $f=78.35$~MHz and $f=75.50$~MHz. For the NMR measurements on the plate-like shaped TbV$_6$Sn$_6$ single crystals, multiple crystals were used with $c$-direction aligned. Field was applied either parallel or perpendicular to the crystaliline $c$-axis. A loosely packed powder sample crushed from the single crystals was also used for comparison, in which the large FM anisotropic $c$-axis would align under a strong magnetic field. NMR measurements are performed at multiple temperatures below $T = 10$~K.  Above 10 K, the nuclear spin-spin relaxation rate $1/T_2$ increases rapidly, leading to a strong reduction of the NMR signals, making the measurements difficult.

{\it DFT+U calculations}. Crystal-field (CEF) parameters are calculated as a function of the aspect ratio $\tfrac{c}{a}$ while keeping the experimentally adopted volume constant.
At each $\tfrac{c}{a}$ value, the CEF parameters are obtained using the total energy mapping method within the constrained DFT+$U$ framework~\cite{ke2025a,han2024proximity}.
The magnetocrystalline anisotropy energy profiles $E(\theta,\phi)$ are determined from total energy calculations as a function of the spin quantization direction.
It is essential to enforce convergence at each spin orientation to the designated $4f$ state within this procedure~\cite{lee2025ncm}.
The validity of this total energy mapping method rests on two recent demonstrations of the success of DFT-based approaches to rare-earth magnetism.
First, excellent agreement between experimental and theoretical $4f$ anisotropy has been achieved across dozens of rare-earth-containing compounds, provided the Hund’s rule ground state is enforced~\cite{PhysRevX.12.021043,PhysRevB.108.045132,PhysRevB.106.115139,xu2024jacs,han2024proximity,gazzah2025a,lee2025ncm}.
Second, the combination of DFT with crystal field theory has recently been validated~\cite{ke2025a}; in the present work, we also confirmed that the extracted CEF parameters can perfectly reproduce the $E(\theta,\phi)$ profiles across all $\tfrac{c}{a}$ values.

Self-consistent DFT+$U$ calculations, including spin-orbit coupling (SOC), were performed to evaluate the total energy $E(\theta,\phi)$ as a function of spin quantization directions, characterized by the polar and azimuthal angles $\theta$ and $\phi$, using the full-potential linearized augmented plane wave (FP-LAPW) method as implemented in \textsc{Wien2k}~\cite{blaha2018book}.
A large value of $U=10$ eV was employed to push the $4f$ states further below the Fermi level~\cite{anisimov1993prb}, and spin-orbit coupling was included using the second-variation approach~\cite{koelling1977jpcs,li1990prb}.
At each spin direction, we ensured that the constrained DFT+$U$ calculations converged to the Hund's rule ground state of Tb$^{3+}$, consistent with experimental results, thereby avoiding the otherwise unphysical DFT+$U$ ground states that arise primarily from the orbital dependence of self-interaction errors~\cite{lee2025ncm}.
Further details of the constrained DFT+$U$ calculations of MAE can be found in Ref.~[\onlinecite{lee2025ncm}].

\begin{acknowledgments}
\section {Acknowledgments}
Inelastic neutron scattering data analysis (T.H., D.J., R.J.M.), X-ray diffraction (A.S.), nuclear magnetic resonance measurements (Q.-P.D., Y.F.), analytical theory (R.D.M., R.F.), and density-functional theory (Y.L., L.K.) calculations performed at the Ames National Laboratory are supported by the U.S. Department of Energy, Office of Science, Basic Energy Sciences, Materials Science and Engineering. Preparation of sample specimens and their characterization (T.J.S., P.C.) is supported by the Center for the Advancement of Topological Semimetals (CATS), an Energy Frontier Research Center funded by the  U.S.\ Department of Energy (DOE) Office of Science (SC), Office of Basic Energy Sciences (BES), through the Ames National Laboratory. Ames National Laboratory is operated for the USDOE by Iowa State University under Contract No. DE-AC02-07CH11358. This research used resources at the Spallation Neutron Source and High Flux Isotope Reactor, DOE Office of Science User Facilities operated by the Oak Ridge National Laboratory. The neutron beam time was allocated to CNCS on proposal number IPTS-29896 and IPTS-33326, POWGEN on proposal number IPTS-31418, and CTAX on proposal number IPTS-32293. Dilatometry measurements (M.L.) acknowledge support from the LDRD program at Los Alamos National Laboratory. A portion of this work was performed at the National High Magnetic Field Laboratory, which is supported by the National Science Foundation Cooperative Agreement No. DMR-2128556* and the State of Florida and the U.S. Department of Energy.
\end{acknowledgments}

% \begin{figure}[]      
%     \centering 
% 	\includegraphics[width=1.\linewidth]{TbV6Sn6_symm_breaking.png}
% 	\caption{\textbf{Effect of symmetry-breaking on spin-flip excitations.} (a) Effect of time-reversal symmetry breaking on Tb spin-flip excitations. (b) Effect of crystallographic symmetry breaking on Tb spin-flip excitations.}
%  \label{fig:splitting}
% \end{figure}

%\bibliographystyle{apsrev4-1}
\bibliography{splitting}
\raggedright

\clearpage
\end{document}

\setcounter{equation}{0}
	\setcounter{figure}{0}
	\setcounter{table}{0}
	\setcounter{page}{1}
	\setcounter{section}{0}
	\makeatletter
    
	\renewcommand{\theequation}{S\arabic{equation}}
	\renewcommand{\thefigure}{S\arabic{figure}}
	\renewcommand{\thetable}{S\arabic{table}}
 	\renewcommand{\thesection}{S\arabic{section}}
	\renewcommand{\bibnumfmt}[1]{[S#1]}
	\renewcommand{\citenumfont}[1]{S#1}

\title{Supplemental Information - Hidden symmetry-breaking in a kagome Ising ferromagnet}
  \author{Tianxiong~Han}
	\affiliation{Ames National Laboratory, U.S. DOE, Iowa State University, Ames, Iowa 50011, USA}
	\affiliation{Department of Physics and Astronomy, Iowa State University, Ames, Iowa 50011, USA}
    \author{Tyler J. Slade}
	\affiliation{Ames National Laboratory, U.S. DOE, Iowa State University, Ames, Iowa 50011, USA}
    \author{Liqin Ke}
	\affiliation{Ames National Laboratory, U.S. DOE, Iowa State University, Ames, Iowa 50011, USA}
    \affiliation{Department of Materials Science and Engineering, University of Virginia, Charlottesville, VA 22904}
    \author{Qing-Ping Ding}
    \affiliation{Ames National Laboratory, U.S. DOE, Iowa State University, Ames, Iowa 50011, USA}
    \author{Minseong Lee}
        \affiliation{ National High Magnetic Field Laboratory, Los Alamos National Laboratory, Los Alamos, New Mexico 87545, USA}
    \author{R. D. McKenzie}
        \affiliation{Ames National Laboratory, U.S. DOE, Iowa State University, Ames, Iowa 50011, USA}
        \affiliation{Department of Physics and Astronomy, Iowa State University, Ames, Iowa 50011, USA}
    \author{Bing~Li}
	\affiliation{Ames National Laboratory, U.S. DOE, Iowa State University, Ames, Iowa 50011, USA}
        \affiliation{Oak Ridge National Laboratory, Oak Ridge, TN, 37831, USA}
    \author{Dhurba R. Jaishi}
	\affiliation{Ames National Laboratory, U.S. DOE, Iowa State University, Ames, Iowa 50011, USA}
	\affiliation{Department of Physics and Astronomy, Iowa State University, Ames, Iowa 50011, USA}
    \author{Yongbin Lee}
	\affiliation{Ames National Laboratory, U.S. DOE, Iowa State University, Ames, Iowa 50011, USA}
    \author{D. M. Pajerowski}
        \affiliation{Oak Ridge National Laboratory, Oak Ridge, TN, 37831, USA}
    \author{Qiang Zhang}
        \affiliation{Oak Ridge National Laboratory, Oak Ridge, TN, 37831, USA}
    \author{Tao Hong}
        \affiliation{Oak Ridge National Laboratory, Oak Ridge, TN, 37831, USA}
    \author{P. C. Canfield}
	\affiliation{Ames National Laboratory, U.S. DOE, Iowa State University, Ames, Iowa 50011, USA}
	\affiliation{Department of Physics and Astronomy, Iowa State University, Ames, Iowa 50011, USA}
    \author{Y. Furukawa}
	\affiliation{Ames National Laboratory, U.S. DOE, Iowa State University, Ames, Iowa 50011, USA}
	\affiliation{Department of Physics and Astronomy, Iowa State University, Ames, Iowa 50011, USA}
    \author{K. Thirunavukkuarasu}
	\affiliation{Florida A\&M University, Tallahassee, FL 32307 USA}
    \author{Aashish Sapkota}
	\affiliation{Ames National Laboratory, U.S. DOE, Iowa State University, Ames, Iowa 50011, USA}
    \author{Rebecca Flint}
	\affiliation{Ames National Laboratory, U.S. DOE, Iowa State University, Ames, Iowa 50011, USA}
	\affiliation{Department of Physics and Astronomy, Iowa State University, Ames, Iowa 50011, USA}
    \author{ R. J. McQueeney}
	\affiliation{Ames National Laboratory, U.S. DOE, Iowa State University, Ames, Iowa 50011, USA}
	\affiliation{Department of Physics and Astronomy, Iowa State University, Ames, Iowa 50011, USA}
\date{\today}

\begin{abstract}
%TbV$_6$Sn$_6$ hosts a V kagome metal combined with strong Ising-like ferromagnetic order within the Tb triangular layer. Deep in the ferromagnetic state, the Ising doublet is split by the molecular field and we expect that the magnetic dynamics consist of a single dispersionless spin-flip excitation. Using inelastic neutron scattering, we rather find two sharp excitations that suggest symmetry breaking.  
\end{abstract}

\maketitle

\section{X-Ray Diffraction}
Table \ref{tbl:XRD} show results of refinements obtained with ShelxL of the single-crystal x-ray diffraction data for TbV$_6$Sn$_6$ at temperatures $T=300$~K, $100$~K, and $80$~K. Measurements were made with a Rigaku XtaLab Synergy-S diffractometer. The changes of the composition in final refinement are negligible ($\leq 0.5\%$).

\section{Powder Neutron Diffraction Refinement}
\begin{figure}[]      
    \centering 
	\includegraphics[width=0.95\linewidth]{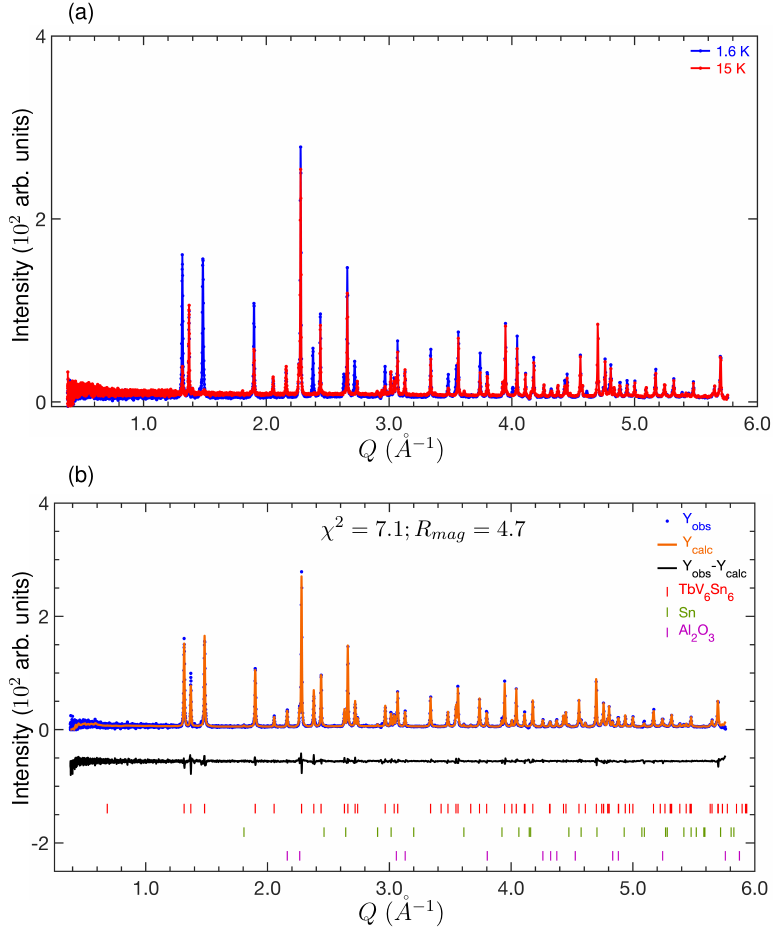}
	\caption{(a). Time-of-flight neutron powder data for TbV$_6$Sn$_6$ measured at $15$ and $1.6K$~K, illustrating the appearance of additional magnetic intensities at $1.6$~K. (b) Rietveld refinement of the $1.6$~K powder data with magnetic symmetry $P6/mm'm'$ and $\chi^2=6.2\%$ and $R_{mag}=3.9$. Blue dots and orange solid lines correspond to the data and calculated patterns, respectively. The black line is the difference between the data and the calculation. Additionally, red, green, and magenta vertical lines correspond to the allowed Bragg reflections for TbV$_6$Sn$_6$ and impurities listed in the figures, respectively.}
 \label{fig:NPD}
\end{figure}

Consistent with the previous reports of the FM ordering ofthe Tb moments, Fig.\ref{fig:NPD} shows the additional magnetic intensities at $\textbf{k}=(0,0,0)$ at $1.6$~K powder diffraction patterns. Additional peaks other than the one corresponding to $P6/mmm$ crystal symmetry of TbV$_6$Sn$_6$ were identified to be from other impurity phases, such as Sn($9\%$) and Al$_2$O$_3$ ($1\%$). For the magnetic Rietveld refinement, maximal magnetic symmetries corresponding to $P6/mmm$ crystal symmetry and propagation vector $\textbf{k}=(0,0,0)$ were calculated by the magnetic symmetry analysis tools using the Bilbao Crystallographic Server \cite{aroyo2006bilbao1,aroyo2006bilbao2,aroyo2006bilbao3,perez2015symmetry}. In our magnetic refinements, we found that magnetic symmetry $P6/mm'm'$ best fits the data, as shown in Fig.\ref{fig:NPD} (b). Magnetic symmetry $P6/mm'm'$ is basically a FM magnetic structure with moment along the $c$-axis. The refined magnetic moment for Tb is $8.6(4)~\mu_B$ at $T = 1.6$~K.

\begin{table}
\caption{Refinements of single-crystal x-ray data for TbV$_6$Sn$_6$.}
\begin{tabular}{c c c c}
\hline\hline
 &TbV$_{6}$Sn$_{6}$ \\
\hline 
Temperature & $295$~K (RT) & $100$~K & $80$~K \\
Space group & P6/mmm &P6/mmm& P6/mmm\\
Lattice constant $a$(\AA) & 5.5244(2) & $5.5138(2)$&$5.5133(2)$ \\
Lattice constant $c$(\AA) & 9.1905(3) & $9.1718(3)$&$9.1691(3)$ \\
$\alpha$ & 90 & 90& 90\\
$\beta$ & 90 & 90& 90\\
$\gamma$ & 120 & 120& 120\\
Volume (\AA$^3$) &242.907(2) &$241.483(2)$ &$241.368(2)$ \\
$Z$ &1 &1 &1 \\
Collected $\Theta_{min}$  &3.360 &$3.367$&$3.367$\\
Collected$\Theta_{max}$  &31.936 & $31.715$& $31.723$ \\
Reflection collected &6534 & 6624&6584 \\
Independent reflections &379 &375 &375 \\
Reflections ($I>2\sigma(I)$) &371 &369 &370 \\
Goodness-of-fit&1.242 &1.162 & 1.116\\
$R_1$ [$I\geq2\sigma(I)$] &0.0122 &0.0102 &$0.0100$ \\
$wR_2$ [$I\geq2\sigma(I)$] &0.0277 &0.0235 &$0.0227$  \\
$R_1$ [all data] &0.0126 &0.0105 &$0.0102$ \\
$wR_2$ [all data] &0.0278 &0.0236 &$0.0228$  \\
Largest peak/ hole &1.060/-1.771 &1.975/-1.206 &$2.319$ / $-0.991$ \\

\hline\hline
\end{tabular}
\label{tbl:XRD}
\end{table}

\section{NMR measurements}
\begin{figure}[]      
    \centering 
	\includegraphics[width=0.95\linewidth]{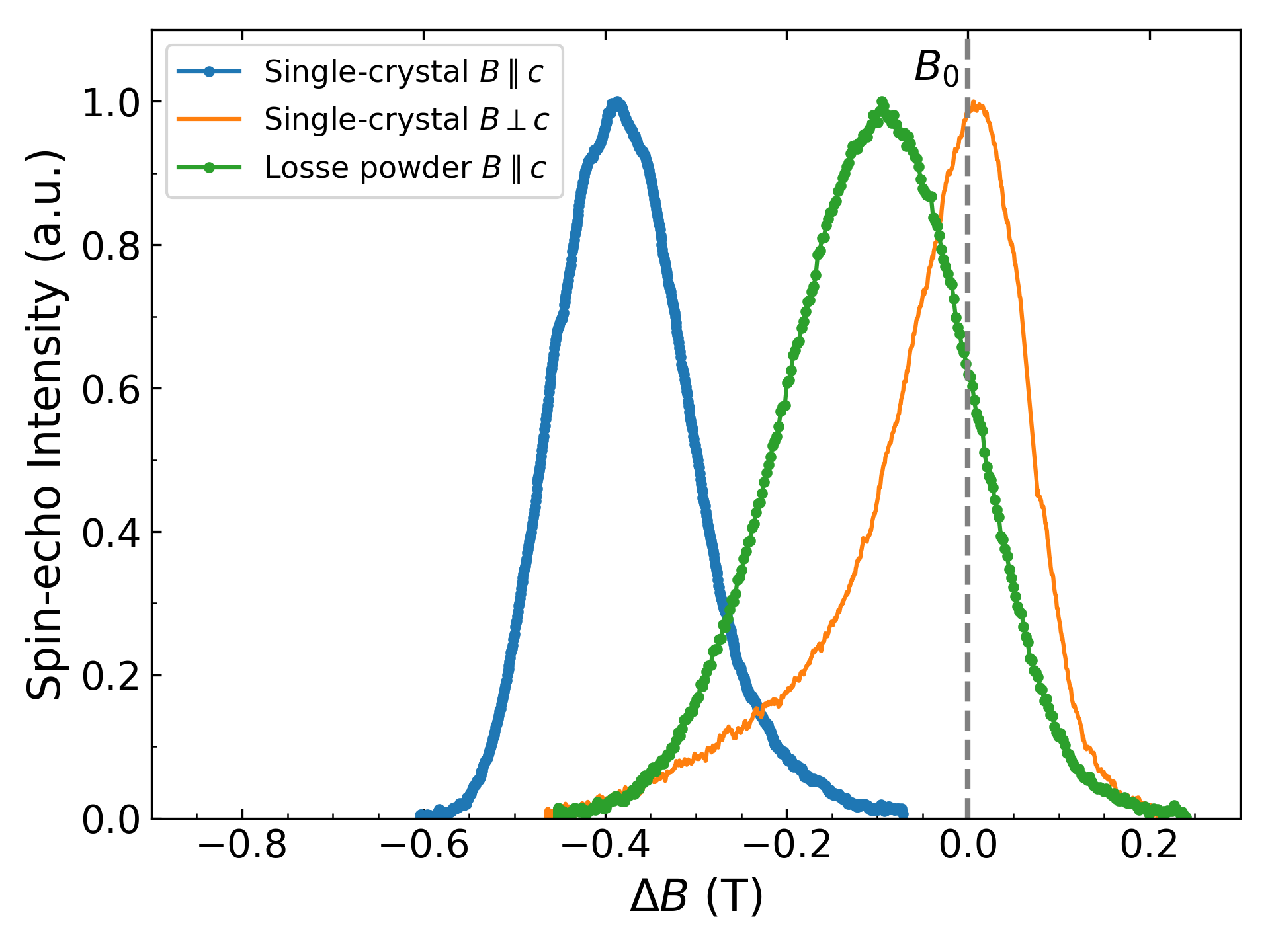}
	\caption{NMR measurement on single crystal and powder samples at $T = 1.6$~K. The intensity is plotted versus the differences of fields $\Delta B$ between the applied field and the Larmor field $B_0=7.0$~T.}
 \label{fig:NMR}
\end{figure}

The nuclear magnetic resonance (NMR) measurements of $^{51}$V nuclei were conducted on a single crystal and loosely packed powder using a laboratory-built phase-coherent spin-echo pulse spectrometer by sweeping the magnetic field $B$ at fixed resonance frequencies.
% The resonance frequencies $f=78.35$~MHz are equivalent to the Larmor field $B_0=7$~T on V sites along $c$-direction, and the resonance frequencies $f=75.5$~MHz are equivalent to the Larmor field $B_0=6.74$~T on V sites perpendicular to $c$-direction. Broad oscillations are consistent with nuclear quadrupole splitting with $\nu_Q = 0.47$ mHz and an anisotropy of $\eta = 0.86$.
Fig.\ref{fig:NMR} shows the NMR spectra measured in the magnetically ordered state \cite{nuQ}, where the horizontal axis $\Delta \vec B$ is the difference between the Larmor field $\vec B_0$ and the applied field $\vec B$.  For measurements above $10$~K without Tb ordering, the nuclear spin-spin relaxation rate $1/T_2$ increases rapidly. This leads to a strong reduction of the NMR signals, making the measurements difficult.

% The difference from the applied field $\vec B$ and the Larmor field $\vec B_0$, is labeled as $\Delta \vec B=\vec B_0-\vec B$. 
The shift of the NMR spectra $\Delta B$ is due to an internal field at the V site originating from the hyperfine field $\vec B_{\text{hf}}$, a classical dipolar field from Tb ordered moments $\vec B_\text{dip}$, and a macroscopic field $\vec B_\text{macro}$. The hyperfine term includes $\vec B_{\text{trans}}$ and $\vec B_{\text{3d}}$, which refer to the transferred hyperfine field between Tb ordered moments and V nuclear spins, and electronic-nuclear spin coupling on the V itself, respectively. The macroscopic field $\vec B_\text{macro}$ includes demagnetization field $\vec B_{\text{dem}}=-\mu_0N \vec M$ and the Lorentz field $\vec B_{\text{L}}=\mu_0 \vec M/3$, where $N$ is the demagnetization factor $N\approx1$ for $H \parallel c$, $N\approx 0$ for $H\parallel ab$ in the plate-like shaped single crystal with the $c$-axis prependicular to the plane, and $N=1/3$ in powder sample. Fig.~\ref{fig:NMR} shows that there is no shift for $H\perp c$, , indicating that all internal fields are parallel to $c$ (we now drop the vector labels for $B$).

We observe a shift $\Delta B = -0.39 (1)$~T determined from the peak position of the spectrum in the single crystal for $H\parallel c$ and $\Delta B=-0.10(1)$~T in the field-aligned powder sample. As the difference originates from the demagnetization term, we estimate $B_{\text{dem}} = -0.44(1)$~T and $B_{\text{L}} = 0.15(1)$~T for $M||c$. These values are in good agreement with the calculated value in the following section. The dipolar field term $B_{\text{dip}} = 0.13$ T is calculated by a lattice sum with $9.0\mu_B/$Tb aligning along the $c$-axis. Hence, the hyperfine field is $ B_\text{hf}\approx B_{\text{trans}}+ B_{\text{3d}} = -0.23(1)$ T. For clarity and ease of interpertation, all contributing local field components are listed in Table \ref{tbl:NMR}.  

Estimating each value of $B_\text{trans}$ and $B_\text{3d}$ in $B_\text{hf}$ is difficult from the present NMR results. Therefore, we here refer to the hyperfine coupling constant $A$ of the V ions in the similar rare-earth bearing compound YbV$_6$Sn$_6$ \cite{park2025} where $A$ along the $c$ axis has been reported to be a positive value of $0.0105$T$/\mu_\text{B}$. This indicates that the sum $B_\text{trans}$ and $B_\text{3d}$ are positive in YbV$_6$Sn$_6$, differing from the negative value of $-0.23(1)$~T (or $A=-0.026(1)$~T$/\mu_\text{B}$) in TvV$_6$Sn$_6$. Since the sign of $B_\text{trans}$ is expected to be the same in two compounds with the same structure, the negative sign in TbV$_6$Sn$_6$ should originate not from $B_\text{trans}$ but from the negative contributions in $B_\text{3d}$. Here we attribute it to the negative core-polarization field from V-$3d$ electrons. Using a typical value of the hyperfine coupling constant for $3d$ electron core-polarization ($-10.5$~T$/\mu_\text{B}$\cite{Freeman1965}) and the observed hyperfine field of $-0.23(1)$~T, the possible V magnetic moment are tentatively estimated to be $0.022(1)\mu_\text{B}$.

% In YbV$_6$Sn$_6$ , the transferred hyperfine coupling constant along the $c$-axis is $A_{trans} = 0.0105~{\rm T}/\mu_B$ \cite{park2025}.  This coupling is expected to be similar in Tb$V_6$Sn$_6$, which suggests that $|B_{3d}| > 0.238$~T.  From this, we can use a typical hyperfine coupling constant for V $3d$ electron core-polarization ($A_{3d} = -10.5$~T$/\mu_B$\cite{Freeman1965})to estimate that a V magnetic moment has a magnitude $\gtrsim 0.02\mu_B$. 

\begin{table}
\caption{Local field (T) on V site from NMR measurements. The positive or negative sign represents parallel or antiparallel to the applied field $B$. The resonance frequency is$f=78.35$~MHz, corresponding to Larmor field $B_0=7.0$~T.}
\begin{tabular}{@{} ccccccc @{}}
\hline\hline
Sample shape& $B$ &\multicolumn{2}{c}{$B_\text{hf}$}&$B_\text{dip}$&\multicolumn{2}{c}{$B_\text{macro}$}\\
\cmidrule(l){3-4} \cmidrule(l){6-7}
    &   &$B_\text{trans}$&$B_\text{3d}$&    &$B_\text{dem}$&$B_\text{L}$\\
\hline
Plate crystal ($N=1$) &7.39(1)&\multicolumn{2}{c}{-0.23(1)}&0.13(1)&-0.44(1)&0.15(1)\\

Powder ($N=1/3$)  &7.10(1)&\multicolumn{2}{c}{-0.23(1)}&0.13(1)&-0.15(1)&0.15(1)\\
\hline\hline
\end{tabular}
\label{tbl:NMR}
\end{table}

% Further NMR studies at low magnetic fields or zero magnetic field would provide more detailed information, especially on AFM ordering proposed in the main text. However, such low magnetic field NMR measurements are challenging because of the shortening of $T_2$.

\section{Raman measurements}
\begin{figure}[]      
    \centering 
	\includegraphics[width=0.95\linewidth]{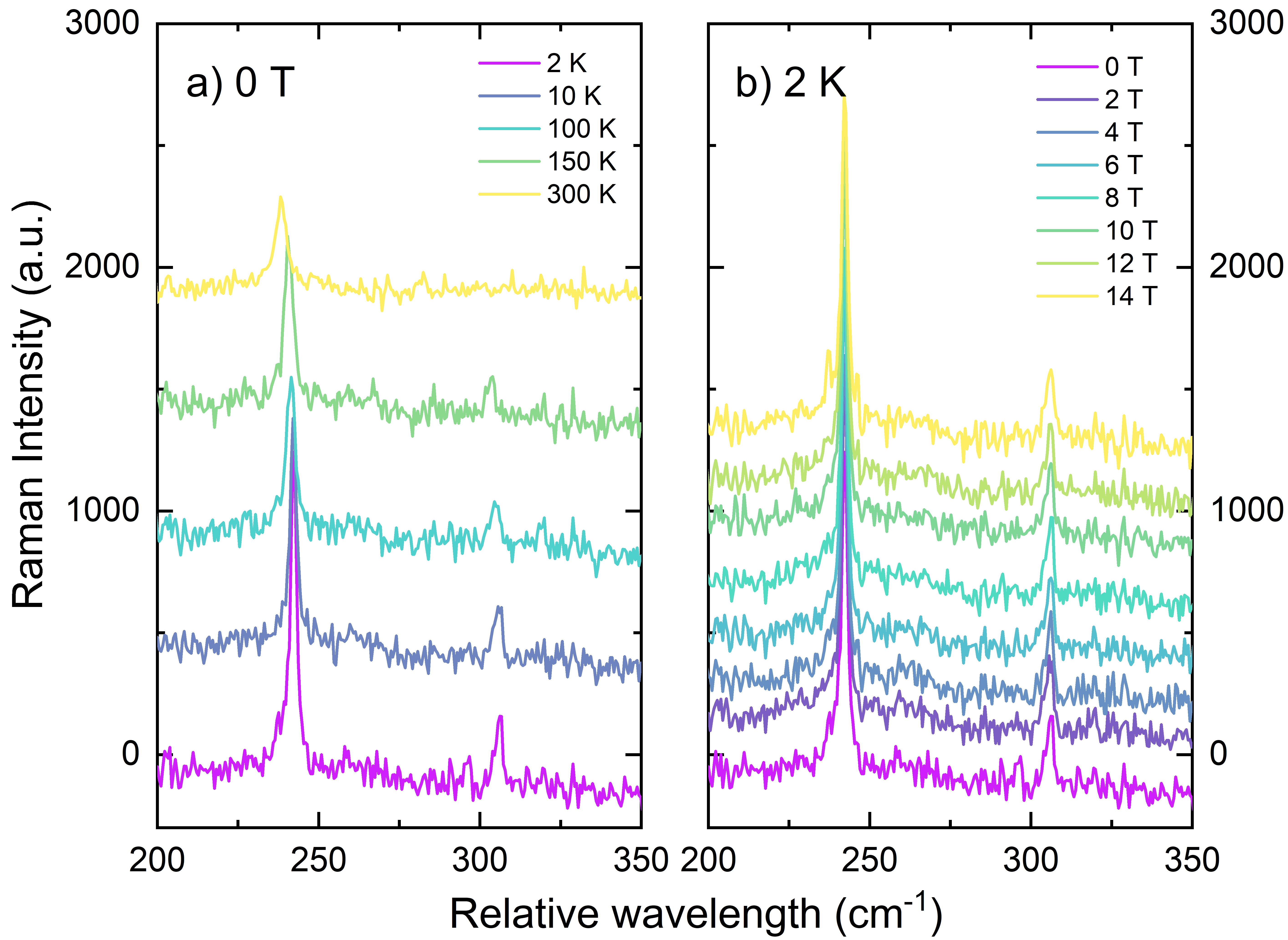}
	\caption{Raman scattering in (a) zero-field as a function of temperature and (b) 2 K as a function of field applied along the $c$-direction (Faraday geometry).}
 \label{fig:raman}
\end{figure}

Raman measurements on TbV$_6$Sn$_6$ were performed using a custom-built free beam optical setup coupled to a PPMS magnet, shown in Fig.\ref{fig:raman}. A laser with $532$~nm wavelength was used for excitation, and back-scattered light in the Faraday geometry was sent to a Princeton Raman spectrometer with liquid nitrogen cooled CCD detector. The incident laser power of $3$~mW was chosen to obtain reliable spectra without heating the sample. The Raman spectra was collected at selected temperatures between $300$~K and $2$~K. The Raman spectra at magnetic field from $0$~T to $14$~T were obtained at the lowest measured temperature $2$~K.

Fig.\ref{fig:raman} shows (a) temperature dependent Raman spectra in the sbsence of magnetic field and (b) Raman spectra with varying magnetic fields between $0$~T and $14$~T at $2$~K. The vertical offset were added in the spectra for clarity. As evident from the Fig.\ref{fig:raman}, the TbV$_6$Sn$_6$ crystal does not show an obvious change in the energy of the phonon modes with decreasing temperature and increasing magnetic field. This suggests that there is an absence of any phase transitions or modifications in the local symmetry of the atoms involved.

\section{Powder Neutron diffraction}

\subsection{Ferromagnetic order parameter}
\begin{figure}[]      
    \centering 
	\includegraphics[width=0.8\linewidth]{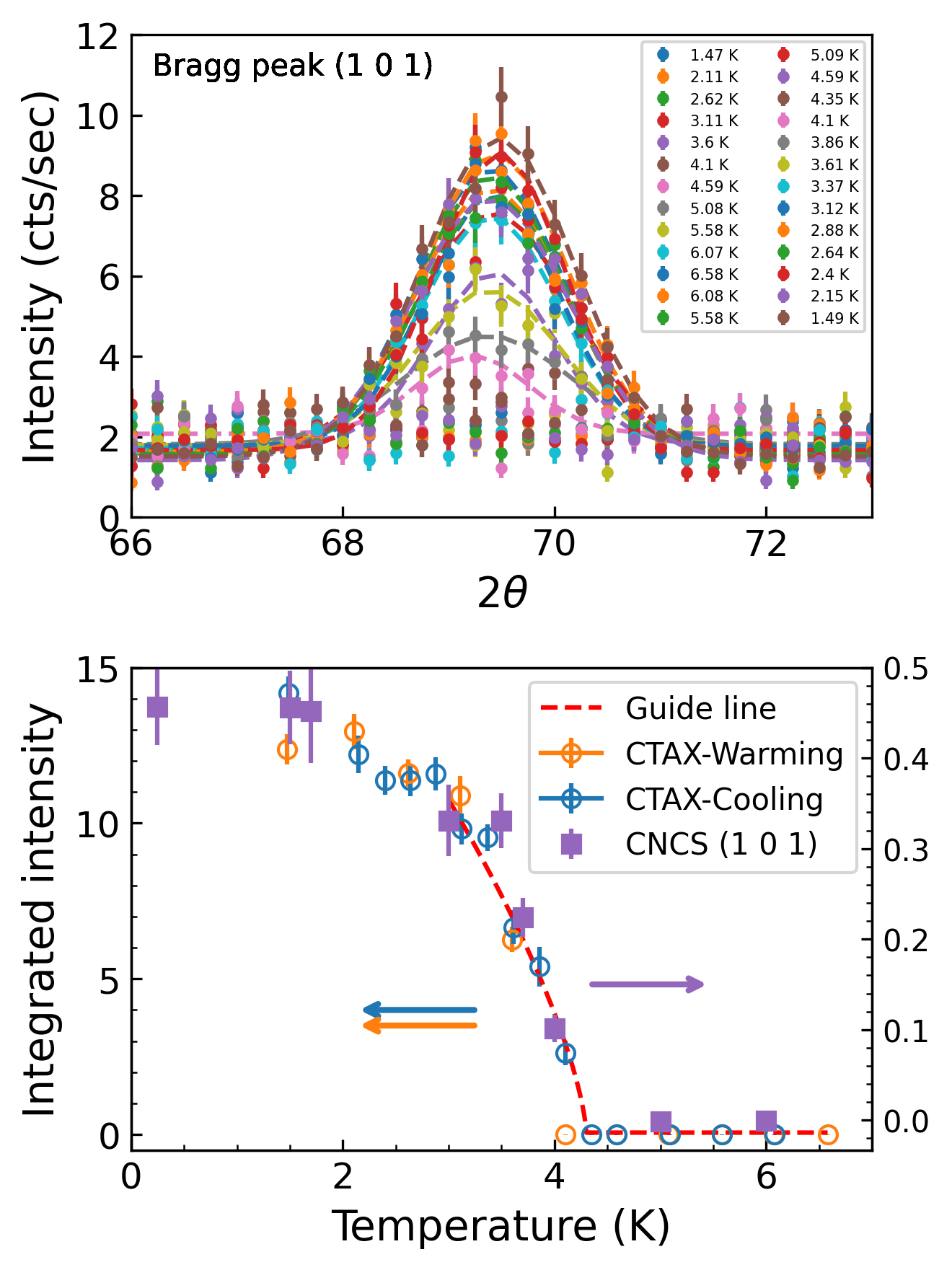}
	\caption{(a) Data from longitudinal scans across the (1,0,1) Bragg peak at various temperatures and the fitting to Gaussian line shapes are shown as dashed lines. (b) The temperature dependence of the integrated (1 0 1) Bragg peak intensities with error bars. The dashed line is a guide for the eyes using $T_{C} = 4.3 (1)$~K.}
 \label{fig:op}
\end{figure}
 Figure \ref{fig:op} shows the temperature dependence of the intensity of $(1, 0, 1)$ Bragg peak measured at CTAX, ORNL with $E_i = 3.5$~meV. These data indicate the temperature dependence of the magnetic order parameter. Fits of Gaussian line shapes to magnetic Bragg peaks showed no significant shifts in the peaks’ centers nor significant changes to their FWHMs below the transition temperature. The estimated $T_C=4.4$ K is consistent with the previous reported results \cite{PhysRevMaterials.6.104202, PhysRevB.106.115139}. The integrated intensity of the peak $(1~0~1)$ from CNCS powder measurement is plotted in Figure \ref{fig:op} (b) as well, showing consistent behavior with the CTAX results.
 \begin{figure}[]      
    \centering 
	\includegraphics[width=0.95\linewidth]{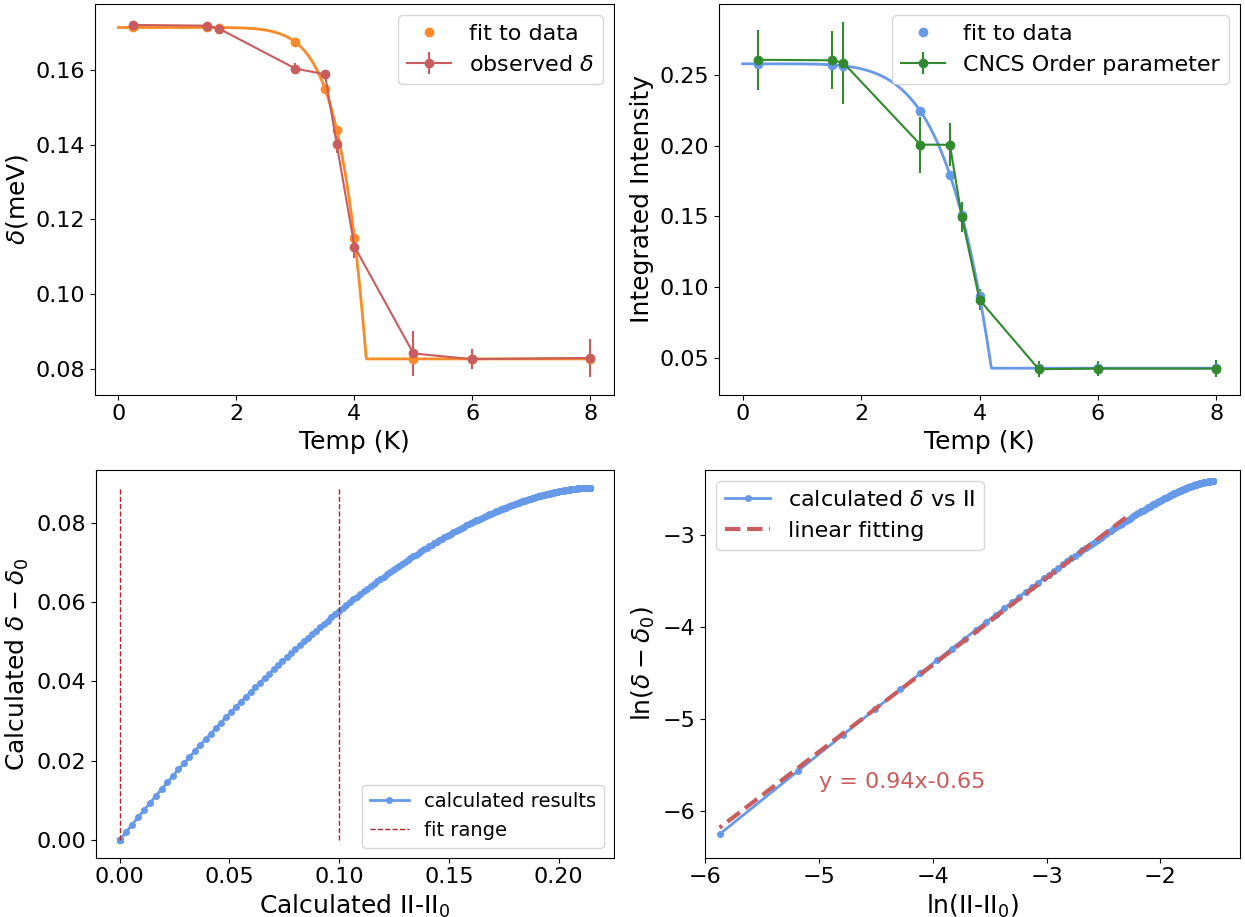}
	\caption{(a) Averaged $\delta$ in the main text and fitted versus temperatures. (b) The temperature dependence of the integrated (1~0~0) Bragg peak intensities with error bars. (c). Calculated $\delta$ versus integrated intensities after subtracting the constants above $T_\text{C}$. The vertical lines indicate the range picked close to $T_\text{C}$ for fitting in the (d) panel. (d). Logarithm of the $\delta$ vs integrated intensities in the range picked, with the linear fitting with slope $k=0.94$.}
 \label{fig:delta_vs_II}
\end{figure}

The splitting $\delta$, as well as the integrated intensities on Bragg peak $(1~0~0)$ measured at CNCS, are plotted and fitted versus temperatures, as shown in Fig.\ref{fig:delta_vs_II}. To interpret the relation between $\delta$ and magnetization $M = \sqrt{\text{II}-\text{II}_0}$, we fit a linear relationship with $\ln{(\delta-\delta_0)}= k \ln{(\text{II}-\text{II}_0)+c}$ near $T_\text{C}$. The fitted slope $k=0.94$ indicates the splitting $\delta\propto M^2$.

 \subsection{Short-range magnetic order}
\begin{figure}[]      
    \centering 
	\includegraphics[width=0.75\linewidth]{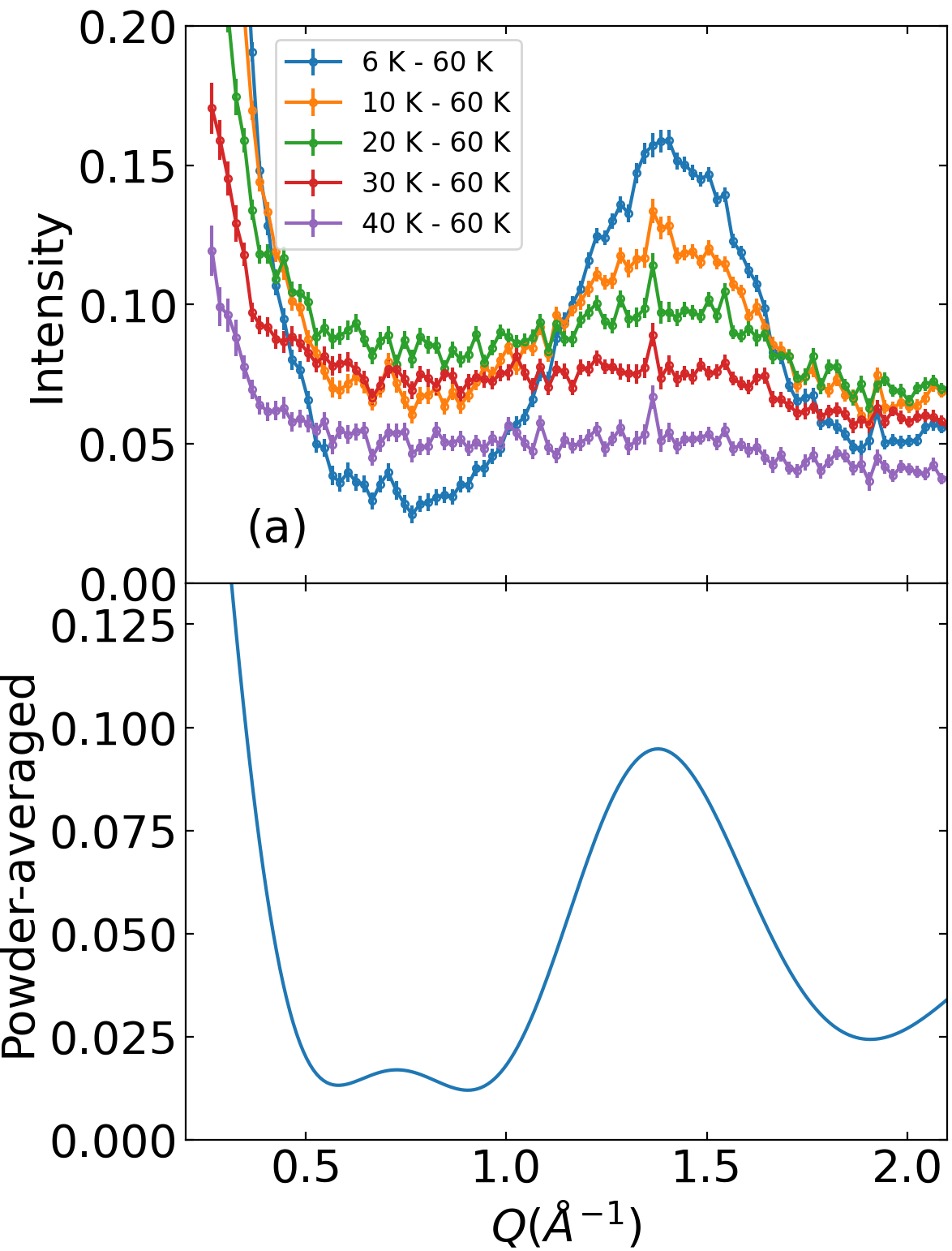}
	\caption{(a) Short-range magnetic correlations obtained from powder inelastic neutron scattering measurements at CNCS. The data were integrated over an energy range $\Delta\hbar\omega\leq0.15$~meV and  the $T=60$~K dataset was subtracted. (b) Calculated powder-averaged structure factor for Tb$^{3+}$ spin cluster on triangular lattice with ion spacing $a=5.51\AA$.}
 \label{fig:sro}
\end{figure}

Short-range magnetic correlations above the ferromagnetic ordering temperature  $T_\text{C} = 4.4$~K were revealed by powder inelastic neutron scattering measurements at CNCS, as shown in Figure \ref{fig:sro}(a). To isolate the quasi-elastic magnetic signal, the intensity was integrated over an energy range $\Delta \hbar\omega \leq 0.15$~meV and data collected at $T = 60$~K were used as a paramagnetic background for subtraction.
The resulting spectrum exhibits broad, temperature-dependent features centered at the positions corresponding to FM Bragg peaks. To interpret this behavior, we calculated the magnetic structure factor for an FM cluster containing six nearest-neighbor Tb$^{3+}$ ions in a triangle lattice, as shown in Figure \ref{fig:sro}(b). The correlation function is written as 
$$I\propto F(Q)^2\sum_{i,j}\langle S_iS_j\rangle e^{i{\bf Q}\cdot ({\bf R}_i-{\bf R}_j)}(1-\frac{Q_z^2}{Q^2})$$ 
where $Q$ is the momentum transfer, $F(Q)$ is the magnetic form factor, and ${\bf R}_i$ is the atomic position. The powder-averaged intensity is calculated by averaging over all directions of ${\bf Q}$.
The calculated profile captures the key features observed in the experimental data, supporting a picture of short-range FM correlations persisting above $T_\text{C}$.

\section{Single-crystal inelastic neutron scattering} 

\subsection{Transverse mode dispersion} 
The main CEF transition from $\ket{6}\rightarrow\ket{5}$ is a strong transverse excitation that can be used to estimate key magnetic interactions. For a simple triangular ferromagnet with nearest-neighbor intralayer ($\mathcal{J}$) and interlayer coupling ($\mathcal{J}_z$), the dispersion in terms of the total angular momentum $J$ can be written as
\begin{align}
\label{eqn:spinwave}
    \hbar\omega & = 2JD + 2J\mathcal{J}_z(1-\cos 2\pi L) +  \nonumber \\
    & 2J\mathcal{J}[3-\cos2\pi(H-K)-\cos2\pi(H+K)-\cos4\pi K]
\end{align}
where $(H,0,0)$ and $(-K,2K,0)$ are orthogonal vectors in the hexagonal plane.  $D$ is the single-ion anisotropy that can be obtained from the CEF parameters according to the relation $D=-K_1/J^2$ where $K_1 = -3J^{(2)}B_2^0 - 40J^{(4)}B_4^0 - 168J^{(6)}B_6^0$ is the magnetic anisotropy constant. Our CEF parameters estimated at 0.25 K ($B_2^0 =$ -0.112, $B_4^0=$-5.41E-4, $B_6^0=$2.31E-6, and $B_6^6=$2.30E-6 meV) find a uniaxial anisotropy parameter $D=-0.641$ meV.
\begin{figure}[]      
    \centering 
	\includegraphics[width=1\linewidth]{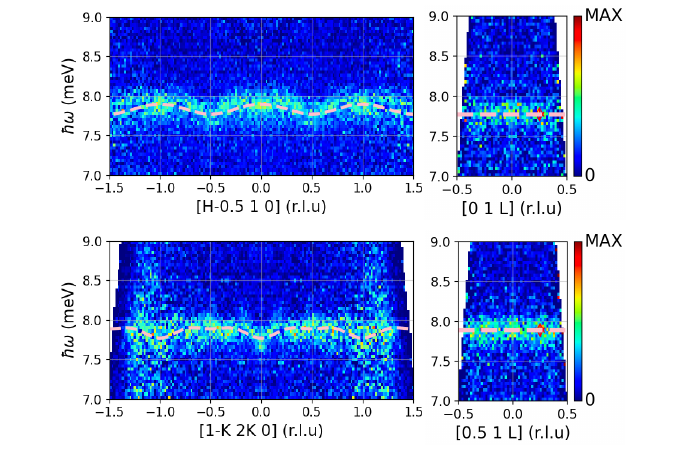}
	\caption{Spin wave dispersion along [H-0.5 1 0], [1-K 2K 0], [0 1 L]and [0.5 1 L] directions. The pink dashed line is the fitted dispersion relation from linear spin wave theory and the red arrow indicates the instrument resolution.}
 \label{fig:spinwave}
\end{figure}

We fit the energy spectra from single-crystal INS measurements shown in Fig.~\ref{fig:spinwave} to Eq.~\ref{eqn:spinwave}, with fitted interaction and anisotropy interaction being $\mathcal{J}=-2.4(2)$~$\mu$eV, $D=-0.65(1)$~meV, and $|\mathcal{J}_z| < 0.0015$~meV. The calculated dispersion from linear spin-wave theory is plotted in the pink dashed line over the dispersion data. 

\subsection{Estimation of dipolar interactions}
The dipole-dipole interactions between Tb moments are given by
\begin{align}
\mathcal{H}_{dip} = \frac{\mathcal{P}}{2} \sum_{i \neq j} \sum_{\mu \nu} 
\mathcal{L}_{ij}^{\mu \nu} J_i^{\mu} J_j^{\nu},
\end{align}
where
\begin{align}
\mathcal{L}_{ij}^{\mu \nu} = \frac{1}{r_{ij}^3}
\biggr(\delta_{\mu \nu}-\frac{3r_{ij}^{\mu}r_{ij}^{\nu}}{r_{ij}^2}\biggr)
\label{eqn:dipole}
\end{align}
and $\mathcal{P}= \mu_0 (g_J\mu_B)^2/(4\pi)$.  Considering the strong uniaxial anisotropy of Tb ions, we estimate the contribution of the dipolar interactions to the local field on a  Tb site for long-range order with wavevector ${\bf k}$ and moments aligned perpendicular to the triangular layer. In a mean-field approximation, the dipolar energy is
\begin{equation}
E_0({\bf k}) = \frac{1}{2}\mathcal{P}J_z^2 L^{zz}({\bf k})
\end{equation}
where $L^{zz}({\bf k}) = \sum_{j}\mathcal{L}^{zz}_{0j}\cos({\bf k}\cdot {\bf r}_{0j})$.
This energy can also be represented as a local magnetic field $B_0^z$ acting on the Tb moment $m_z = g_J\mu_{\rm B}J_z$.
\begin{equation}
B_0^z= -\mathcal{P}J_z{L}^{zz}({\bf k})/(g_J\mu_{\rm B}) = -\frac{2E_0(\bf k)}{m_z}
\end{equation}
For FM order, this local field must also be corrected for demagnetization effects given by 
\begin{equation}
    B_{local}^z=B_{Lorentz}-\mu_0NM+B_0^z
\end{equation}
where $N$ is the demagnetization factor and $B_{Lorentz}=\mu_0M/3$ is the net field inside a spherical cavity of a uniformly magnetized volume.  For TbV$_6$Sn$_6$, $M = \frac{9\mu_{\rm B}}{\sqrt{3}a^2c/2}$ from which we obtain $B_{Lorentz} = 0.1438$ T. For a needle-shaped FM domain with $N=0$, the local field is $B_{local}^z = B_{Lorentz}+B_0^z$.

Table \ref{tbl:Dipole} shows select dipolar energies and local fields for the simple hexagonal Tb sublattice with $c/a = 1.66$.  Stripe-type AFM order [${\bf k}=(1/2,1/2,0)$] is favored over FM order [${\bf k}=(0,0,0)$] within the triangular layer. This clearly establishes that dipolar coupling competes with a dominant RKKY-type FM exchange within the layer. From calculations of the dipolar coupling using Eqn.~\ref{eqn:dipole}, we estimate that $\mathcal{J}_{dip}^{zz}/\mathcal{J}_{RKKY} \approx -0.2$ for nearest-neighbors within the layer.

The dispersionless character of excitations along $(0,0,L)$ shown in Fig.~\ref{fig:spinwave} indicate very small interlayer magnetic interactions and establish the quasi-2D character of TbV$_6$Sn$_6$. The interlayer interactions leading to long-range FM order could be dipolar in character and we estimate a FM dipolar coupling of $\mathcal{J}_z'\approx -.00031$ meV for interlayer nearest-neighbors. However, the long period magnetic incommensurability along $L$, as reported for GdV$_6$Sn$_6$ \cite{PhysRevB.108.035134}, may also indicate a role for weak intralayer RKKY interactions.

\begin{table}
\caption{Dipolar sums for TbV$_6$Sn$_6$ with moments aligned perpendicular to the triangular layer (in meV). A local field correction appropriate for a long needle-shaped domain is made in the FM case (${\bf k}=0$).}
\begin{tabular}{c|c|c}
\hline\hline
{\bf k} & $E_0({\bf k})$ (meV) & $B_{local}^z$ (T) \\
\hline 
(0,0,0) & +0.0666 & -0.112 \\
(1/2,1/2,0) & -0.0268 & 0.103 \\
(0,0,1/2) & +0.1410 & -0.541 \\
\hline\hline
\end{tabular}
\label{tbl:Dipole}
\end{table}

%\subsection{Magnetoelastic interactions}
% Fig.~\ref{fig:main_CEF_temperature} shows the temperature dependence of the main CEF transition energy ($\Delta$) and also its peak width. Below $T_C$, the transition energy decreases by $\approx 1.7\%$.  The decrease could arise from magnetoleastic interactions. 

% Since the magnetic anisotropy energy scale dominates the exchange, we consider that magnetoelastic interactions originate from single-ion anisotropy with form
% \begin{equation}
%     \mathcal{H}^{me}_i = \sum_l \lambda_{\alpha}^{l0} \epsilon_{\alpha}\mathcal{O}_l^0({\bf J}_i) + \lambda_{\alpha}^{66}\epsilon_{\alpha}\mathcal{O}_6^6({\bf J}_i)
% \end{equation}
% where $\epsilon_l$ are symmetry-conserving lattice strains and $\lambda_{\alpha}^{lm}$ are the variation of the CEF parameters under the strain field \cite{jensen1991rare}.   DFT and point-charge calculations of $\lambda_{\alpha}^{lm}$ in Fig.~3(b) allow an estimate that $\Delta c/ c\sim -0.3\%$ for full ferromagnetic saturation. However, a much smaller spontaneous magnetostriction of $\sim 0.003\%$ is observed in Fig.~3(e) of the main manuscript.  The reason for this discrepancy is not known.

% \begin{figure}[]      
%     \centering 
% 	\includegraphics[width=0.7\linewidth]{main_CEF_temp_dependence.png}
% 	\caption{Temperature dependence of (a) the main CEF excitation energy of TbV$_6$Sn$_6$ and (b) its peakwidth.  The solid line in (a) corresponds to a descrease of $M^2$.}
%  \label{fig:main_CEF_temperature}
% \end{figure}

\bibliography{splitting_SI}
\raggedright

\clearpage